\documentclass[fleqn,usenatbib]{mnras}

\usepackage[T1]{fontenc}

\DeclareRobustCommand{\VAN}[3]{#2}
\let\VANthebibliography\thebibliography
\def\thebibliography{\DeclareRobustCommand{\VAN}[3]{##3}\VANthebibliography}

\usepackage{graphicx}	
\usepackage{amsmath}	
\usepackage{amssymb}	
\usepackage{color}
\usepackage{booktabs}
\usepackage{breakcites}
\usepackage[table]{xcolor}
\usepackage{empheq}
\usepackage{newtxtext,newtxmath}

\mathchardef\mhyphen="2D

\newcommand{\di}{\mathrm{d}}
\newcommand{\bfx}{\mathbf{x}}

\newcommand{\vlos}{{v}_{\rm los}}
\newcommand{\NSD}{\mathrm{NSD}}
\newcommand{\BAR}{\mathrm{BAR}}

\newcommand{\degree}{\ensuremath{^\circ}}

\newcommand{\bfv}{\mathbf{v}}
\newcommand{\bfJ}{\mathbf{J}}
\newcommand{\pc}{\,{\rm pc}}
\newcommand{\kpc}{\,{\rm kpc}}

\newcommand{\Gyr}{\,{\rm Gyr}}
\newcommand{\kms}{\,{\rm km\, s^{-1}}}

\newcommand{\e}{\mathrm{e}}
\newcommand{\Msun}{\, \rm M_\odot}
\newcommand{\Msunyr}{\, \rm M_\odot\, yr^{-1}}
\newcommand{\masyr}{\, \rm mas\, yr^{-1}}

\usepackage{xcolor}
\definecolor{mypink}{rgb}{0.458, 0.188, 0.478}
\definecolor{myblue}{rgb}{0.0, 0.3, 0.8}

\title[Self-consistent modelling of the NSD]{Self-consistent modelling of the Milky Way's Nuclear Stellar Disc}

\author[Sormani et al.]{%
Mattia C. Sormani,$^{1}$\thanks{E-mail: mattia.sormani@uni-heidelberg.de}
Jason L. Sanders,$^{2,3}$ 
Tobias K. Fritz,$^{4}$
Leigh C. Smith,$^{3}$ \newauthor
Ortwin Gerhard,$^{5}$
Rainer Sch\"{o}del,$^{6}$ 
John Magorrian,$^{7}$ 
Nadine Neumayer,$^{8}$ \newauthor 
Francisco Nogueras-Lara,$^{8}$  
Anja Feldmeier-Krause,$^{8}$
Alessandra Mastrobuono-Battisti,$^{9,10}$ \newauthor
Mathias Schultheis,$^{11}$
Banafsheh Shahzamanian,$^{6}$ 
Eugene Vasiliev,$^{3}$ 
Ralf S. Klessen,$^{1,12}$ \newauthor
Philip Lucas,$^{13}$
and Dante Minniti$^{14,15}$
\\
$^1$ Universit\"{a}t Heidelberg, Zentrum f\"{u}r Astronomie, Institut f\"{u}r theoretische Astrophysik, Albert-Ueberle-Str. 2, 69120 Heidelberg, Germany \\
$^2$ Department of Physics and Astronomy, University College London, London WC1E 6BT, UK \\
$^3$ Institute of Astronomy, University of Cambridge, Madingley Rd, Cambridge, CB3 0HA, UK \\
$^4$ Department of Astronomy, University of Virginia, Charlottesville, 530 McCormick Road, VA 22904-4325, USA \\
$^5$ Max-Planck-Institut fur Extraterrestrische Physik, Gie{\ss}enbachstra{\ss}e, D-85748 Garching, Germany \\
$^6$ Instituto de Astrofisica de Andalucia (CSIC), Glorieta de la Astronomia s/n, 18008 Granada, Spain \\
$^7$ Rudolf Peierls Centre for Theoretical Physics, Clarendon Laboratory, Parks Road, Oxford OX1 3PU \\
$^8$ Max Planck Institute for Astronomy. K\"{o}nigstuhl 17. D-69 117 Heidelberg, Germany \\
$^{9}$ Department of Astronomy and Theoretical Physics, Lund Observatory, Box 43, SE-221 00, Lund, Sweden \\
$^{10}$ GEPI, Observatoire de Paris, PSL Research University, CNRS, Place Jules Janssen, 92190 Meudon, France \\
$^{11}$ Universit\'e C\^ote d’Azur, Observatoire de la C\^ote d'Azur, Laboratoire Lagrange, CNRS, Blvd de l'Observatoire, F-06304 Nice, France\\
$^{12}$ Universit\"at Heidelberg, Interdisziplin\"ares Zentrum f\"ur Wissenschaftliches Rechnen, Im Neuenheimer Feld 205, D-69120 Heidelberg, Germany \\
$^{13}$ Centre for Astrophysics Research, University of Hertfordshire, College Lane, Hatfield, AL10 9AB, UK \\
$^{14}$ Depto. de Cs. F\'isicas, Facultad de Ciencias Exactas, Universidad Andr\'es Bello, Av. Fern\'andez Concha 700, Las Condes, Santiago, Chile \\
$^{15}$ Vatican Observatory, V00120 Vatican City State, Italy
}

\pubyear{2021}

\begin{document}
\label{firstpage}
\pagerange{\pageref{firstpage}--\pageref{lastpage}}
\maketitle

\begin{abstract}
The Nuclear Stellar Disc (NSD) is a flattened high-density stellar structure that dominates the gravitational field of the Milky Way at Galactocentric radius $30\lesssim R\lesssim 300\pc$. We construct axisymmetric self-consistent equilibrium dynamical models of the NSD in which the distribution function is an analytic function of the action variables. We fit the models to the normalised kinematic distributions (line-of-sight velocities + VIRAC2 proper motions) of stars in the NSD survey of Fritz et al., taking the foreground contamination due to the Galactic Bar explicitly into account using an $N$-body model. The posterior marginalised probability distributions give a total mass of $M_{\rm NSD} = 10.5^{+1.1}_{-1.0} \times10^8 \,\Msun$, roughly exponential radial and vertical scale-lengths of $R_{\rm disc} = 88.6^{+9.2}_{-6.9} \pc$ and $H_{\rm disc}=28.4^{+5.5}_{-5.5} \pc$ respectively, and a velocity dispersion $\sigma \simeq 70\kms$ that decreases with radius. We find that the assumption that the NSD is axisymmetric provides a good representation of the data. We quantify contamination from the Galactic Bar in the sample, which is substantial in most observed fields. Our models provide the full 6D (position+velocity) distribution function of the NSD, which can be used to generate predictions for future surveys. We make the models publicly available as part of the software package \textsc{Agama}.
\end{abstract}



\section{Introduction} \label{sec:introduction}

The centre of the Milky Way harbours a Nuclear Stellar Disc (NSD), a flattened high-density stellar structure that dominates the gravitational field at Galactocentric radius $30\lesssim R \lesssim 300\pc$. The NSD is part of the Nuclear Bulge, which can be defined as the region within Galactocentric radius $R\simeq300\pc$ and consists of the NSD, a much more compact and more spherical Nuclear Star Cluster (NSC), the central black hole Sgr~A*, and an accumulation of dense and star-forming gas known as the Central Molecular Zone (CMZ) \citep{Launhardt2002}. The NSD is not isolated but is embedded at the centre of the much larger Galactic Bulge/Bar.

\subsection{Structure of the NSD} \label{sec:structure}

The first comprehensive description of the NSD can be found in \citet{Launhardt2002}, although hints at its existence can be found in previous works \citep{Catchpole1990,Lindqvist1992}. \citet{Launhardt2002} report a radius of $R=230\pm20\pc$ from COBE infrared photometry, a vertical scale-height of $h=45\pm 5\pc$ from warm dust (used as a proxy of the stellar distribution due to the low resolution of the COBE data), and a total mass of $M=1.4 \pm 0.6 \times 10^9 \Msun$. \cite{Nishiyama2013} fit exponentials to star counts in the $H$ and $K$ bands and find a similar scale-height of $h=45\pm 3\pc$. \cite{Schodel2014b} find a somewhat smaller scale-height of $h\simeq30\pc$ using Spitzer/IRAC infrared photometry; however, they do not explicitly study the structural properties of the NSD but they fit it as a background for their study of the NSC. \cite{Gallego-Cano2020} study the NSD using two different datasets (the stellar density map from \citealt{Nishiyama2013} and Spitzer/IRAC $4.5\micron$ images) and find a radial scale-length of $R\simeq90\pc$ and a scale-height of about $h\simeq30\pc$. \cite{Debattista2015,Debattista2018} argue for a much larger NSD with radius of $\simeq 1\kpc$ as an explanation for the presence of high-velocity peaks in the line-of-sight velocity distribution of stars near the Galactic Centre at $4\degree<l<14\degree$, which however can be also explained by stars on elongated orbits in the Galactic Bar \citep{Molloy2015,Aumer2015,Zhou2021}.

The NSD is rotating. The rotation of the NSD has been detected in APOGEE spectroscopic data by \cite{Schonrich2015}, in OH/IR and SiO maser stars by \cite{Lindqvist1992} and \cite{Habing2006}, in ISAAC (VLT) near-infrared integral-field
spectroscopy by \cite{Feldmeier2014}, in classical cepheids by \cite{Matsunaga2015}, in the KMOS spectroscopic survey by \cite{Fritz2021}, and in proper motions parallel to the Galactic plane by \cite{Shahzamanian2021}.

Whether the NSD is axisymmetric is an open question. There have been some suggestions in the literature, based on a longitudinal asymmetry in 2MASS star count maps \citep{Alard2001,Rodriguez-Fernandez2008} and on a change in the orientation of the Bar at small longitudes measured by using Red Clump stars as standard candles \citep{Nishiyama2005,Gonzalez2011}, that the NSD may actually be a non-axisymmetric nuclear bar. Indeed, roughly 30\% of nearby barred galaxies host a secondary nuclear bar \citep{Erwin2011}. However, \citet{Gerhard2012} (see also \citealt{Valenti2016}) have shown that the asymmetry observed in the Milky Way can also be explained by geometric projection effects of the large-scale Bar combined with an axisymmetric NSD. Extinction can also produce an apparent asymmetry in the stellar distribution, since distribution of dust in the CMZ is highly asymmetric, with most of it being located at positive longitude \citep[e.g.][]{Molinari2011,Alonso-Garcia2017,Nogueras-Lara2021b}, consistent with the location of the apparent deficit of stars in the 2MASS maps. This asymmetry is also obvious in the extinction maps in Figure 6 of \cite{Schodel2014b} and in the star counts in Figure 1 of \cite{Nishiyama2013}, which show clearly the presence of dark patches at positive latitudes correlating with the position of dark clouds. Thus, current observational constraints appear consistent with the NSD being an axisymmetric structure, although it cannot be ruled out that it consists of a secondary nuclear bar.

\subsection{Formation and evolution of the NSD} \label{sec:formation_intro}

Nuclear stellar discs are common in the centre of barred spiral galaxies \citep{Pizzella2002,Gadotti2019,Gadotti2020}. They are expected to form as follows, although the details are not completely understood. Interstellar gas is channelled by galactic bars towards the centre along features known as the bar ``dust-lanes'' with typical inflow rates of a few $\Msunyr$ \citep{Regan1997,Laine1999,Elmegreen2009,Shimizu2019,Sormani2019c}. This gas accumulates in the centre where it forms gaseous nuclear rings with typical radii that range from a few tens of pc to a kpc in radius \citep{Comeron2010}. These rings are vigorously star forming, and the nuclear discs are the long-term product of this star formation over secular timescales. 

Consistent with the above picture, \cite{Gadotti2020} find that nuclear stellar discs in nearby galaxies are characterised by near-circular rotation and low velocity dispersions. \cite{Bittner2020} further support the idea that nuclear discs form from star formation in gaseous nuclear rings by showing that nuclear discs are younger, more metal-rich, and show lower $[\alpha/{\rm Fe}]$ enhancements than their immediate surroundings. They also find that nuclear discs exhibit well-defined
radial gradients, with ages and metallicities decreasing with radius.
They interpret these gradients as evidence that nuclear discs grow inside-out, from a series of gaseous rings that grow in radius over time. This inside-out formation scenario is perfectly consistent with results of hydrodynamical simulations that show that the size of nuclear rings increases with the amount of stellar mass in the centre \citep{Athanassoula1992,Seo2019}. This suggests that it is the mass increase of the nuclear disc itself that causes the nuclear ring to grow bigger, so that the next generation of stars forms at a slightly larger radius than the previous one.

The above findings for nearby galaxies are mirrored by similar findings in the Milky Way. The gaseous ring-like structure in the Milky Way is known as the CMZ \citep{Morris1996}. The current bar-driven mass inflow rate onto the CMZ is $\dot{M}=0.8\pm0.6\Msunyr$ \citep{Sormani2019c,Hatchfield2021} and its current star formation rate (SFR) is $\simeq 0.1\Msunyr$ \citep{Barnes2017}.  The NSD and the dense gas ring in the CMZ overlap in radius and have comparable scale-heights \citep{Molinari2011,Henshaw2016,Longmore2017}. Stars in the NSD are kinematically cold and rotate with velocities similar to those of the dense gas in the CMZ \citep{Schonrich2015,Schultheis2021}. Furthermore, stars in the NSD have a metallicity distribution function that is different from those of the NSC and of the Galactic Bulge \citep{Schultheis2021}. These findings support the hypothesis of a strong link between star formation in the CMZ and the formation of the NSD and are fully consistent with the formation picture described above.

Assuming that the NSD forms from star formation in the CMZ, an open question is how is the star formation distributed in time. \cite{Figer2004} determined the star formation history (SFH) in pencil beam fields throughout the NSD and argue for a quasi-continuous SFR over the last $\sim 10\, \Gyr$. Assuming that the SFR in the CMZ has been constant over the last $10\Gyr$ at the current rate of $\simeq 0.1\Msunyr$ gives a total stellar mass of $M\simeq 10^9\Msun$, very similar to the current mass of the NSD. However, this appears to be a mere coincidence since more recently \cite{Nogueras-Lara2020b} used the GALACTICNUCLEUS survey to determine the SFH over a more extended region in the NSD and found evidence for a variable SFR. By modelling the extinction-corrected $K$-band colour-magnitude diagram as a superposition of star formation events at different times, they conclude that $\sim 80 \%$ of the stars in the NSD formed more than $8 \Gyr$ ago, followed by a drop in star formation activity between $1$ and $8\Gyr$ ago, and then by a more recent increased activity in the last Gyr. Assuming that the Milky Way Bar is older than $8 \Gyr$, the SFH determined by \cite{Nogueras-Lara2020b} would be consistent with recent simulations from \cite{Baba2020} that predict that the Bar formation triggers an intense star formation episode that lasts for $\sim 1 \Gyr$ followed by lower amounts of variable star formation during the subsequent gigayears (see in particular their Figure 3). The emerging picture is therefore the following: most of the mass of the NSD formed shortly after the formation of the Bar $>8 \Gyr$ ago. Then, from $8 \Gyr$ ago to the present day, the NSD has grown further at variable rates depending on the rate of SFR in the CMZ, which is regulated by the amount of fresh gas available through the bar-driven inflow and possibly by internal feedback cycles (for discussions on what controls the star formation rate in the CMZ see for example \citealt{Kruijssen2014,Krumholz2017,Armillotta2019,Sormani2020b,Moon2021a,Moon2021b}).

\subsection{Dynamical models}
 
Understanding the structure and dynamics of the NSD using dynamical models is important for a number of reasons. First, nuclear stellar discs have a higher height-to-radius ratio (i.e.\ are puffed up), a shorter dynamical time and a completely different formation history than the better-studied galactic discs. Thus studying them can give us new insight on the kinetic theory and heating mechanisms of stellar discs. Second, constraining the gravitational potential created by the NSD is crucially important to understand the gas flows in the CMZ and the inward transport of gas from the CMZ down to the central black hole \citep{Tress2020a,Li2021}. Third, we need to first study equilibrium models if we want to understand the instabilities that might lead to the formation of inner bars \citep{Erwin2011,deLorenzo-Caceres2019,Bittner2021}. Fourth, having a model of the distribution of stars in 6D (position+velocity) phase space can be useful for a number of applications such as generate predictions for future surveys or inferring the 3D position of highly extincted dark clouds \citep{Zoccali2021,Nogueras-Lara2021a}.

In a recent paper \citet{Sormani2020a} constructed axisymmetric Jeans models of the NSD to constrain its properties. They found a total NSD mass of $M_{\rm NSD}=(6.9\pm 2) \times 10^8 \Msun$, gave an analytical 3D model for its density distribution and constrained the velocity dispersion. However, while Jeans models are useful as a first step in assessing the dynamical properties of a system, they are intrinsically limited since (i) they only rely on moments of the collisionless Boltzmann equation and do not provide a full 6D ($\bfx$,$\bfv$) representation of the system under study; (ii) the Jeans equation of the $n$-th moment involves the $n+2$-th moment, so an ansatz is required to close the hierarchy of equations; (iii) there is no guarantee that the models are physical, i.e. that an underlying non-negative distribution function exists; (iv) it is almost impossible to deal properly with the effects of extinction and selection functions.

In this paper we construct self-consistent axisymmetric equilibrium models of the NSD in which the distribution function is an analytic function of the action variables. These models overcome the shortcomings of the Jeans models and provide the full 6D density distribution in phase space. We fit these models to the spectroscopic NSD survey of \cite{Fritz2021} cross-matched with the VIRAC2 proper motion catalogue from Smith et al.\ (in prep). 

The paper is structured as follows. In Section \ref{sec:data} we describe the observational data. In Section \ref{sec:selection} we derive a selection function that characterises the probability that a star ends up in our sample given the observational selection criteria. In Section \ref{sec:scm} we describe the self-consistent modelling methodology, and in Section \ref{sec:fitting} the fitting procedure. In Section \ref{sec:results} we present our results and in Section \ref{sec:discussion} we discuss them. We sum up in Section \ref{sec:conclusion}.

\section{Observational Data} \label{sec:data}

\begin{figure}

	\includegraphics[width=\columnwidth]{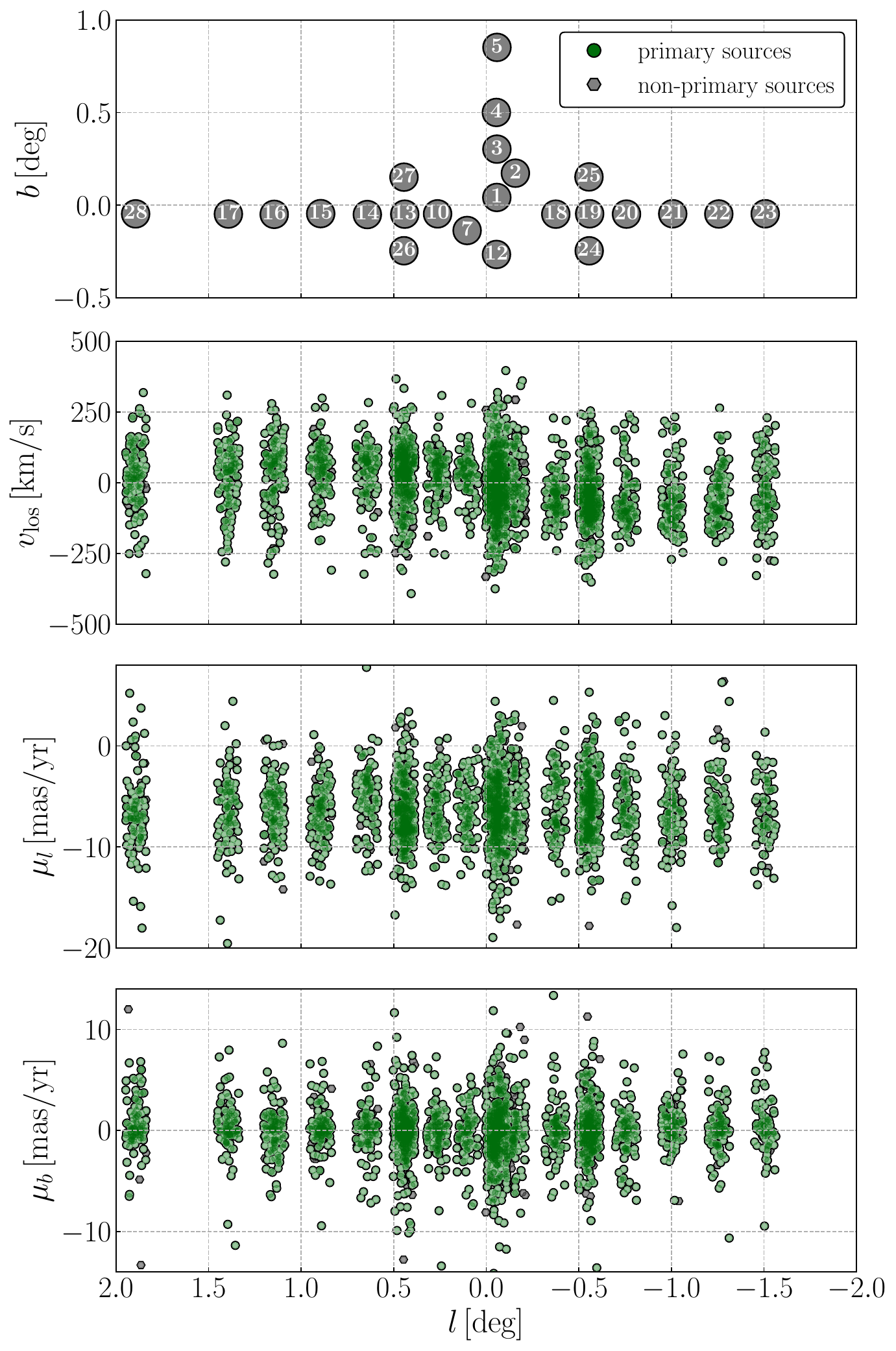}
    
    \caption{The KMOS NSD survey of \citet{Fritz2021} cross-matched with the VIRAC2 reduction of VVV. The top panel shows the fields observed in the survey, numbered according to Table A.1 in \citet{Fritz2021}. Each point in the other three panels represents an individual star. In green and gray are the primary sources and non-primary sources of the survey respectively. $\vlos$ is the line-of-sight velocity. $\mu_l$ and $\mu_b$ are the proper motions in the Galactic longitude and latitude direction respectively.}
    \label{fig:kmos_01}
\end{figure}

We use data from the KMOS spectroscopic survey of \citet{Fritz2021}. This is a dedicated study of the NSD and the innermost Bar/Bulge in the infrared $K$-band,\footnote{In this paper, we use $K$ as a shorthand to denote the $K_s$ band with centre at $2.150\micron$.} containing a total of 3065 stars. The design and strategy of the survey are described in detail in \citet{Fritz2021}. The top panel in Figure~\ref{fig:kmos_01} shows the fields observed in the survey, numbered as in Table A.1 of \citet{Fritz2021}.

\citet{Fritz2021} provide line-of-sight velocities, but they do not provide proper motions. For proper motion data, we cross-matched the KMOS survey with preliminary data from the VIRAC2 photometric and astrometric reduction of the Vista Variables in the Via Lactea survey (VVV) data \citep{Minniti2010}. VVV is a multi-epoch near-infrared survey of the Galactic Bulge and southern Disc with observations in $ZYJHK$ spanning an approximately ten year baseline. \citet{Smith2019} describe the first version of the VVV Infrared Astrometric Catalogue (VIRAC), which use the multi-epoch $K$ aperture photometry from VVV to measure relative proper motions. These are then fixed to the Gaia DR2 absolute reference frame by \citet{Sanders2019}. The 2nd version of VIRAC (VIRAC2, Smith et al., in prep.) improves on the first version by using (i) point-spread-function photometry, (ii) an increased number of $K$ epochs (those of the VVVX temporal extension to VVV) and (iii) a calibration to Gaia DR2 astrometric reference frame at the image level.
The KMOS sources are cross-matched to VIRAC2 within a $0.4\,\mathrm{arcsec}$ radius utilising the proper motions to account for the epoch difference. Only high confidence VIRAC2 sources are considered, i.e. those not flagged as duplicates and with five-parameter astrometric solutions. For stars with $|\mu_{b,\rm err}|<1 \masyr$ the typical scatter between the $J$, $H$ and $K$ photometry from VIRAC2 compared to SIRIUS \citep{Nagayama2003,Nishiyama2006} is $\sim0.2\,\mathrm{mag}$ with offset magnitudes $\lesssim0.06\,\mathrm{mag}$ (not accounting for the different photometric systems). There are three outliers with significantly brighter SIRIUS $K$ magnitudes than VIRAC $K$. Comparison with Spitzer/IRAC $[3.6]$ \citep{Churchwell2009} suggests the SIRIUS measurements are spurious for these sources. 
From the cross-match we obtain proper motions for 2533 out of the 3065 stars in the KMOS survey. Most of the KMOS sources fainter than $K=10.5$ are successfully cross-matched. For stars brighter than $K\approx11$, saturation effects begin appearing in VVV.  Thus, for stars significantly brighter than this limit there are typically an insufficient number of unsaturated observations to obtain an astrometric solution. The mode of the proper motion uncertainty distribution is around $0.3\,\mathrm{mas\,yr}^{-1}$ with a long tail towards higher uncertainties (the 90th percentile is at $1.5\,\mathrm{mas\,yr}^{-1}$). Assuming the stars are located at the Galactic Centre, the mode uncertainty corresponds to a transverse velocity uncertainty of $\sim12\,\mathrm{km\,s}^{-1}$ ($\sim60\,\mathrm{km\,s}^{-1}$ for the 90th percentile). Blending in these crowded regions could produce spurious VIRAC2 proper motions, the impact of which is difficult to assess without higher resolution imaging. However, our sample is limited to bright stars ($K\lesssim13$) for which systematics from blending are expected to be small.

We define the sample used in the fitting procedure (Section \ref{sec:fitting}) as follows. First, we only include stars that are primary sources of the survey (see \citealt{Fritz2021} for definitions). This leaves us with 2805 stars out of the initial 3065. Of these remaining stars, 2803 have line-of-sight velocity and 2316 have proper motions. However, a significant fraction of proper motions have large errors, in particular when we approach $K\simeq 11$ due to the previously described saturation effects. Thus, we choose to keep $\mu_l$ only for stars with $|\mu_{l,\rm err}|<1 \masyr$ and $\mu_b$ only for stars with $|\mu_{b,\rm err}|<1 \masyr$. The final sample used in our fitting procedure contains 2805 stars of which 2803 have $\vlos$, 1908 have $\mu_l$, and 1900 have $\mu_b$. 

\begin{figure*}

	\includegraphics[width=\textwidth]{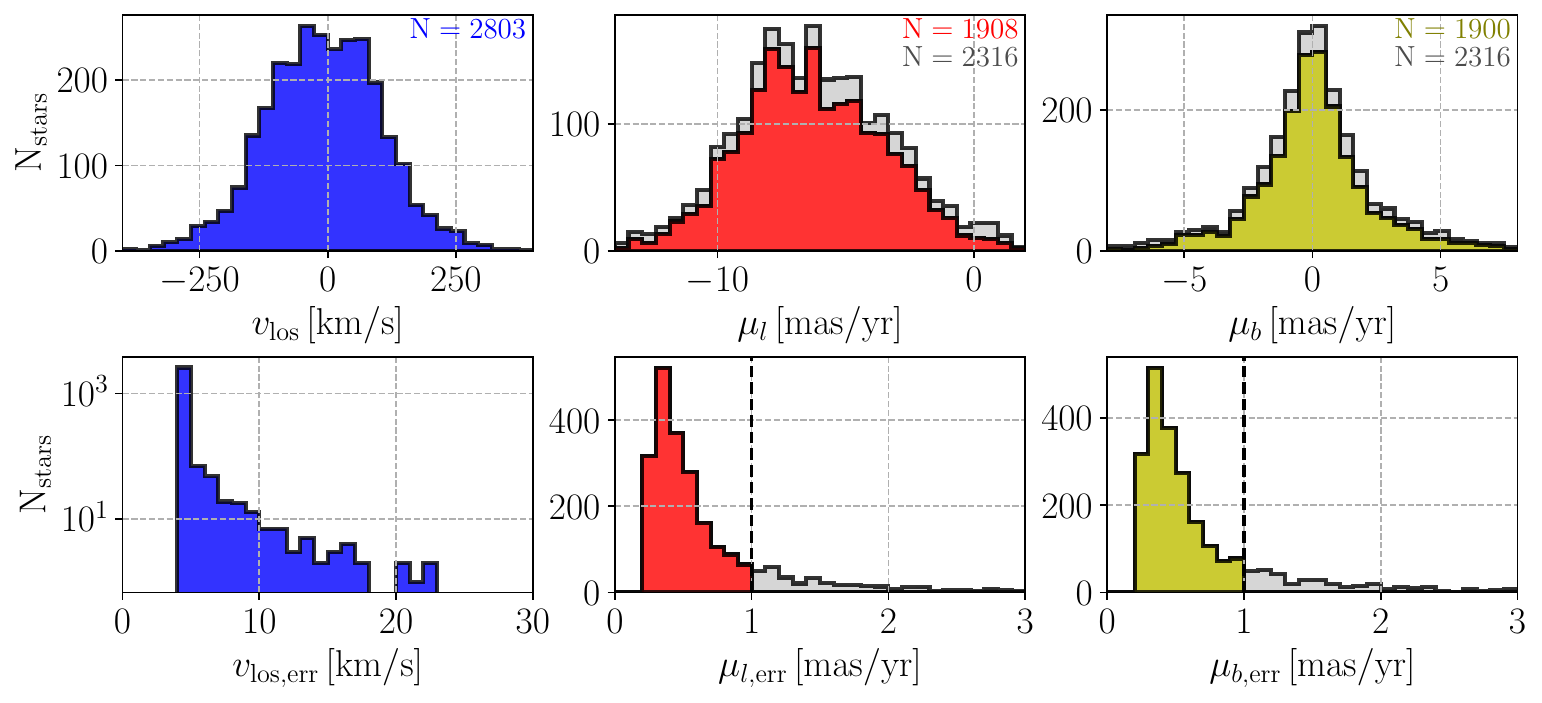}
    
    \caption{\emph{Top row}: histograms of line-of-sight velocities and proper motions in our sample. \emph{Bottom row:} the corresponding observational errors. Numbers annotated in the top panels indicate the total number of stars in each histogram. The vertical black dashed line in the bottom panels indicates the quality cut that we applied on proper motions (see Sect.\ \ref{sec:fitting}). Gray are the histograms before the quality cut, while in colour (red or yellow) are the histograms with the stars remaining after the cut. The distribution of $\mu_l$ proper motions is not centred around $0$ because these are absolute proper motions, not relative, and therefore the central black hole has a finite proper motion of about $\mu_l\simeq6.4 \masyr$ which is due to the orbit of the Sun around the Galactic Centre \citep[e.g.][]{Reid2004}.}
    \label{fig:kmos_02}
\end{figure*}

\begin{figure}

	\includegraphics[width=\columnwidth]{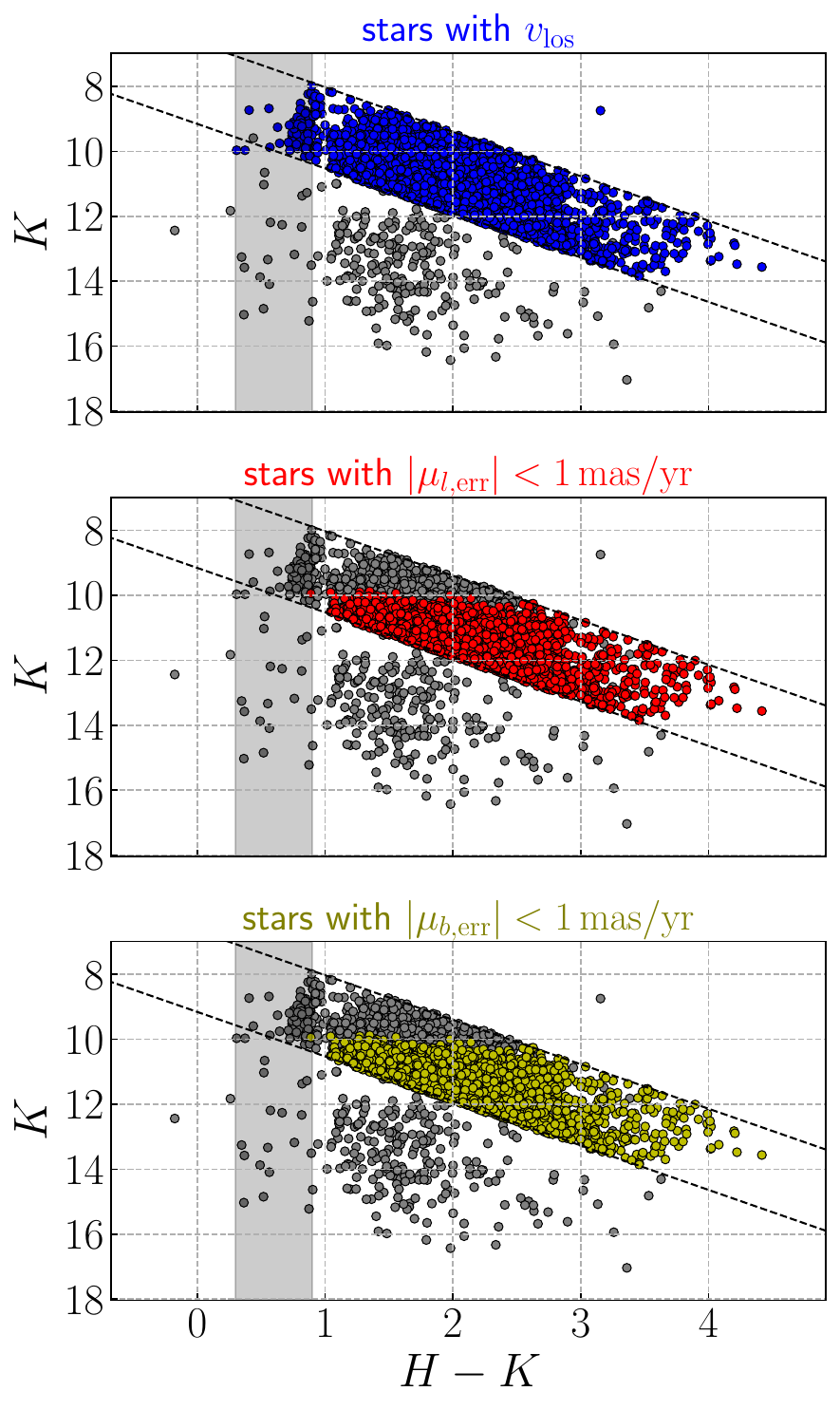}
    
    \caption{Colour-magnitude diagram of stars in our sample. The coloured stars are those included in our sample and used in our fitting procedure (Blue: stars with $\vlos$, Red: stars with $\mu_l$, Yellow: stars with $\mu_b$), while the gray stars are those excluded from our sample. The saturation effect that prevents us from obtaining high-quality proper motions for stars with $K\gtrsim10$ is evident as a magnitude cut in the bottom two panels. The two diagonal black dashed lines $K=1.37(H-K) + 6.6575$ and $K=1.37(H-K) + 9.1575$ represent one of the selection criteria of the \citet{Fritz2021} survey (see Section~\ref{sec:selection}). The gray shaded region shows the range in which the cut $(H-K)_{\rm cut}$ is applied in the survey, which varies by field from 0.3 to 0.9 (see Section~\ref{sec:selection}). The colours in this figure match those in Figure \ref{fig:kmos_02}.}
    \label{fig:kmos_03}
\end{figure}

Figure~\ref{fig:kmos_02} shows the distributions of stars and their errors in our sample, while Figure~\ref{fig:kmos_03} shows their colour-magnitude diagrams.

\section{Selection Function} \label{sec:selection}

\begin{figure}
	\includegraphics[width=\columnwidth]{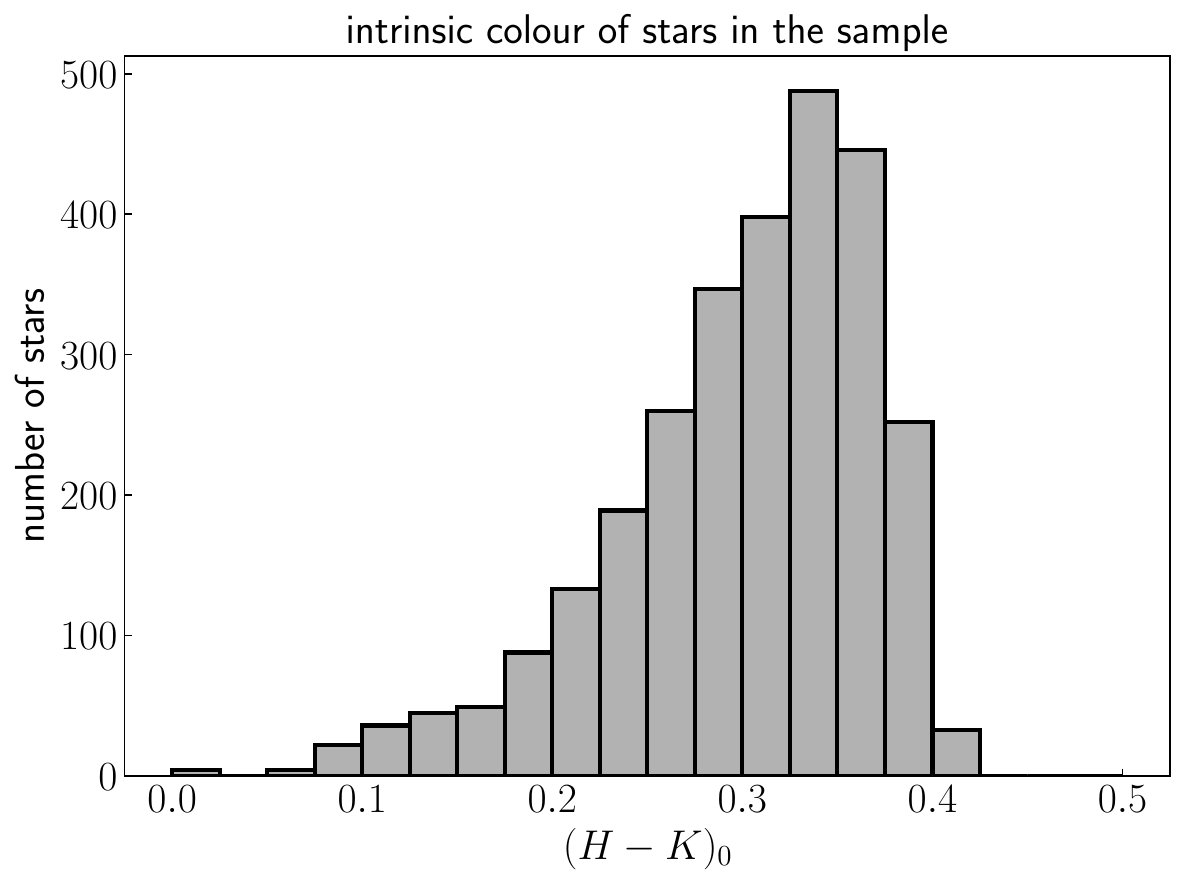}
    \caption{Distribution of intrinsic colour $(H-K)_0$ for stars in our sample. We used the values provided by \citet{Fritz2021} derived from the spectroscopic parameters.}
    \label{fig:kmos_04}
\end{figure}

As is evident from Figure \ref{fig:kmos_03}, the stars in our sample only cover certain parallelogram-shaped areas of the $K\mhyphen(H-K)$ plane. This is mainly the result of the following three selection criteria:
\begin{enumerate}
    \item The survey of \citet{Fritz2021} selects only stars in the region $6.6575<K-1.37 \times (H-K)< 9.1575$ (i.e. between the two diagonal black dashed lines in Figure~\ref{fig:kmos_03}). This corresponds to selecting stars with unextincted apparent magnitude between $7.0<K_0<9.5$ if we assume that (1) the intrinsic colour of the observed stars is $(H-K)_0=0.25$ (typical for red giants, which constitute $>99
    \%$ of all the stars in the survey), and that (2) extinction is related to the colour excess by \citep[see discussion in][]{Fritz2021} \begin{equation} \label{eq:1}
    A_K=1.37 \times [(H-K)-(H-K)_0]\,.\end{equation}
    In reality, stars in the sample do not all have exactly the same $(H-K)_0$ (see Figure \ref{fig:kmos_04}), but this is a small effect, so to a good approximation stars in our sample have unextincted magnitudes in the range $7.0<K_0<9.5$.
    \item The survey contains a blue cut that excludes stars with $(H-K)<(H-K)_{\rm cut}$. The value of $(H-K)_{\rm cut}$ is field-dependent and varies from 0.3 in high latitude fields to 0.9 in the mid plane. Since extinction is related to the $(H-K)$ colour via Equation~\eqref{eq:1} and since stars in our sample are confined to a small range in $(H-K)_0$ (see Figure~\ref{fig:kmos_04}), this is essentially a cut on extinction, hence its primary effect is to remove foreground objects (see also the intersection between the two lines in the left panels in Figure~\ref{fig:sel_01}).
    \item The survey has a magnitude truncation so that only stars with $8<K<14$ are retained. For proper motions, there is an additional effective magnitude cut at $K\simeq 10$ because proper motions with errors less than 1 mas/yr are not available if the stars are too bright. This is visible in the bottom two panels of Figure~\ref{fig:kmos_03}.
\end{enumerate}
The effect of these selection criteria that is of concern to us here is that the probability of observing a star depends on its distance from us. Because intrinsically brighter stars (i.e., with larger absolute magnitude $M_K$) are rarer than fainter stars along the giant branch, and because stars in our sample are limited in apparent magnitude ($7.0\lesssim K_0 \lesssim 9.5$, see above), stars closer to us are preferentially observed than those that are further away. This effect is negligible for stars within the NSD, but is significant for stars belonging to the Galactic Bar/Disc, so we need to model it. We do so by introducing a selection fraction $S_{k,j}(d)$. This is defined as the fraction of stars that end up in our survey given a fixed amount of stellar mass at distance $d$. We normalise this fraction by its value at the Galactic Centre, $S_{k,j}(8.2\kpc)=1$ \citep{GravityCollaboration2019}. So for example, if $S_{k,j}(5\kpc)=2.5$ there will be 2.5 times more stars that end up in our survey from a given stellar population placed at $d=5\kpc$ than if the same population were placed at the Galactic Centre.

The selection fraction is field-dependent (index $j$), because, as mentioned in the second item above, the value of $(H-K)_{\rm cut}$ varies slightly from field to field. The selection fraction also depends on whether we consider line-of-sight velocities or proper motions (index $k$), because only proper motions have the $K\simeq 10$ cut in the colour-magnitude distribution (see third criterion above and Figure~\ref{fig:kmos_03}).

\begin{figure}
	\includegraphics[width=\columnwidth]{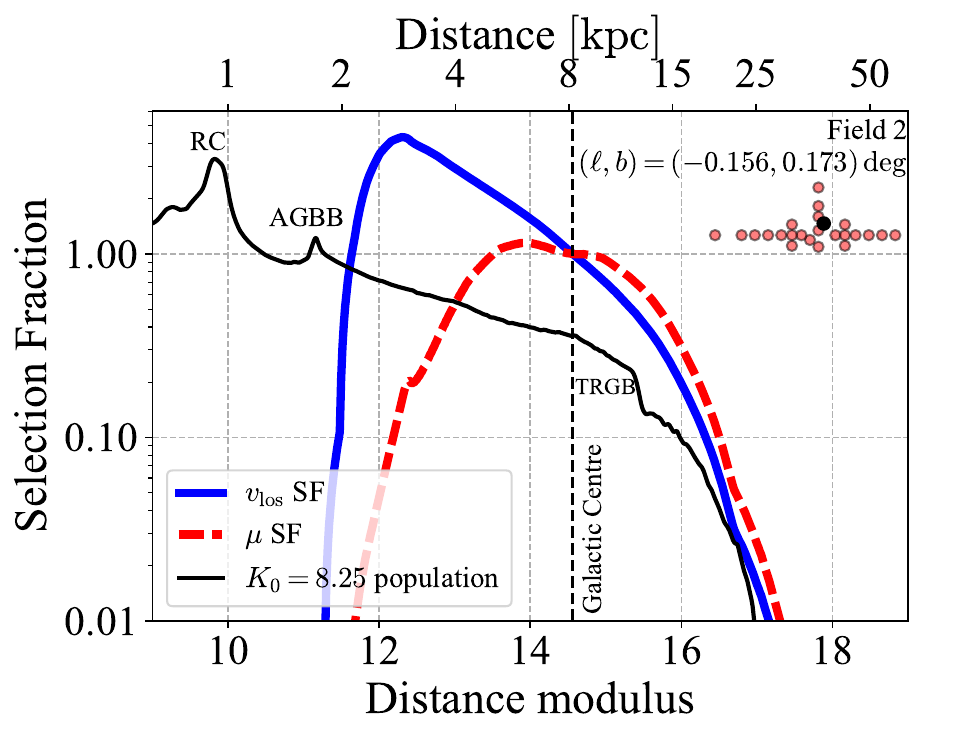}
    \caption{The selection fraction $S_{k,j}$ for an example field (see discussion in Section~\ref{sec:selection}). \emph{Blue solid}: the selection fraction for $\vlos$. In this example field, stars closer to us are preferentially selected up to a sharp drop at a distance of approximately $d\simeq 3\kpc$. \emph{Red dashed}: the selection fraction for the proper motions (the ones for longitude and latitude proper motions are nearly identical). The vertical dashed line indicates the position of the Galactic Centre. The black solid line is the distribution of a population of stars all with unextincted apparent magnitude $K_0=8.25$.
    }
    \label{fig:sel_main}
\end{figure}

We construct the selection fraction $S_{k,j}$ through the following steps:
\begin{enumerate}
    \item We first incorporate the criteria that only stars with $7.0<K_0<9.5$ are selected. We generate stars from a \cite{Kroupa2001} initial mass function (IMF) between a minimum mass of $0.01 \Msun$ and a maximum mass of $100 \Msun$. We then evolve these stars to present day for a range of different ages and metallicities using stellar evolution tables from the PARSEC team \citep{Bressan2012,Marigo2017}. For each age and metallicity, we compute the one-dimensional distributions of stars in absolute magnitudes $M_K$. We normalise these distributions by their total contribution of stellar mass today (stars + remnants which are now black dwarfs, black holes and neutron stars)\footnote{We neglect mass lost due to stellar winds. We have checked that this does not affect the selection fraction significantly.} using the prescription from \cite{Maraston1998}. The reason why we do the normalisation in this way is that we need to translate number of observed stars into the mass of the population they represent (because our models are defined in terms of mass per phase space). We then sum these distributions by weighting by age according to the star formation history in the Galactic Bulge from \citet{Bernard2018} and by metallicity according to the metallicity distribution function from the combined spectroscopic sample studied by \citet{Schultheis2019}. We thus obtain the $M_K$ distribution of stars at present day for a given mass of star formation. We then convert this $M_K$ distribution into a 2D distribution of stars as a function of unextincted $K_0$ and distance $d$ using the relation
    \begin{equation} \label{eq:2}
        K_0 = M_K + 5 \log_{10}\left(\frac{d}{10\pc}\right)\,,
    \end{equation} 
    and for each distance we find the fraction of stars that fall within the range $7<K_0<9.5$. This fraction normalised by its value at $8.2\kpc$ gives us a selection fraction that takes into account the first selection criterion above. The black line in Figure~\ref{fig:sel_main} shows the distribution of stars with $K_0=8.25$ as a function of distance modulus as an illustrative example.
    \item  To incorporate the $(H-K)_{\rm cut}$ we proceed as follows. We take the distribution of intrinsic $(H-K)_0$ in our sample (Figure~\ref{fig:kmos_04}), and assume that the same distribution is valid for the stellar populations generated in the previous step (independently of the distance). Then for each given distance and for each field we get the colour excess $E(H-K)$ and its uncertainty from the 3D extinction map of \citet{Schultheis2014}. We take the intrinsic $(H-K)_0$ distribution, shift it by $E(H-K)$ and broaden it by the uncertainty. In this way we obtain a predicted $(H-K)$ distribution for each distance  and for each field. Finally, we use these distributions to calculate the fraction of stars with $(H-K)>(H-K)_{\rm cut}$ at each distance and in each field. This gives a distance-dependent and field-dependent multiplicative factor that is incorporated into the selection fraction.
    \item Finally, we incorporate the truncation $8<K<14$ and the magnitude cut on stars with good proper motions as follows. We take the 2D distribution of stars as a function of $K_0$ and $d$ calculated at step (i) and convert it into a 2D distribution as a function of $K$ and $d$ using the relation $K=K_0 + A_K$, Equation \eqref{eq:1} and the values of $E(H-K)$ from the \cite{Schultheis2014} 3D map. We then look at the fraction of stars that fall outside the truncation $8<K<14$ as a function of distance and include this as a multiplicative factor in the selection fraction. For the proper motions only, we histogram in $K$ all the stars in our sample in the given field and the subset with good proper motions (see middle panel in Figures~\ref{fig:sel_01}-\ref{fig:sel_04}). We then take the ratio as a function of $K$ as a multiplicative factor that is included in the selection fraction.
\end{enumerate}
In Appendix \ref{sec:appendix_sel} we report the selection fraction obtained for all fields.

\section{Self-consistent model} \label{sec:scm}

The models are made of two components: the NSC and the NSD. The properties of the NSC have already been well constrained in previous work, so we treat it as a fixed external component defined by its axisymmetric density distribution $\rho_{\rm NSC}(R,z)$ (Section \ref{sec:NSC}). We do not have to worry about contamination of stars from the NSC in our sample because the KMOS NSD data of \citet{Fritz2021} avoids the inner few arcmins which are occupied by the NSC. It is not necessary to include the central black hole Sgr~A* in our models since its gravitational potential dominates at $R\lesssim 1\pc$, much smaller than the scales of interest here.

The NSD is assumed to be an axisymmetric, collisionless stellar system in dynamical equilibrium within the gravitational potential created by itself and the NSC. The NSD is defined by an analytic distribution function (DF) $f(\bfJ)$ which is a function of the action integrals (Section \ref{sec:NSD}). 

\subsection{The Nuclear Star Cluster} \label{sec:NSC}

The NSC is a fixed component that simply provides an external contribution to the overall gravitational potential. The mass distribution that generates the potential of the NSC is taken to be the best-fitting axisymmetric model from \citet{Chatzopoulos2015b}:
\begin{align}
\rho_{\rm NSC}(R,z) 	& = \frac{(3-\gamma)M_{\rm NSC}}{4\pi q}\frac{a_0}{a^{\gamma} (a+a_0)^{4-\gamma}} \,,\label{eq:NSC}  
\end{align}
where
\begin{equation}
a(R,z) = \sqrt{R^2 + \frac{z^2}{q^2}} \,,
\end{equation}
and $\gamma=0.71$, $q=0.73$, $a_0=5.9\pc$, and $M_{\rm NSC}=6.1\times10^7 \Msun$. The total cluster mass of this model is significantly higher than the mass given in \cite{Schodel2014b} or in \cite{Feldmeier-Krause2017}, but this does not make a significant difference on our results for the NSD. The NSC has no free parameters.

\subsection{The Nuclear Stellar Disc} \label{sec:NSD}

According to Jeans's theorem, the DF of a steady-state system must depend on the phase-space coordinates only through the integrals of motion \citep[e.g.][]{Binney2008}. A particularly convenient choice is to use the action variables $\bfJ=(J_R,J_\phi,J_z)$ as integrals of motion because these variables also act as the conjugate momenta of the action-angle coordinate system \citep[e.g.][]{Binney2013}. We define the NSD with a quasi-isothermal DF which is a parametrised function of the three action variables \citep{Binney2010,Binney2011,Vasiliev2019}:
\begin{equation} \label{eq:qiso}
f(\bfJ) = \frac{\tilde{\Sigma} \Omega}{2 \pi^2 \kappa^2} \frac{\kappa}{\tilde{\sigma}_r^2} \e^{-\frac{\kappa J_R}{\tilde{\sigma}_r^2}} \frac{\nu}{\tilde{\sigma}_z^2} 	\e^{-\frac{\nu J_z}{\tilde{\sigma}_z^2}}	\times \begin{cases}1 & \text{if } J_\phi\geq0 \\ \e^{\frac{2\Omega J_\phi}{\tilde{\sigma}_r^2}} & \text{if } J_\phi < 0 \end{cases}		\,,
\end{equation}
where 
\begin{align}
\tilde{\Sigma}(R_c) 	 & = \Sigma_0 \e^{-\frac{R_c}{R_{\rm disc}}} \,, \\
\tilde{\sigma_r}^2(R_c) & = \sigma_{r,0}^2 \e^{- \frac{2 R_c}{R_{\sigma,r}}} + \sigma_{\rm min}^2 \,, \label{eq:sigmar}\\
\tilde{\sigma_z}^2(R_c) & = 2 H_{\rm disc}^2 \nu^2(R_c) + \sigma_{\rm min}^2 \,. \label{eq:sigmaz}
\end{align}
Here, $J_R \geq 0$ is the radial action which describes radial oscillations, $J_z \geq 0$ is the vertical action describing oscillations out of the $z=0$ plane and $J_\phi= R v_\phi$ is the azimuthal action which coincides with the conserved $z$-component of the angular momentum, and can have both signs. $R_c(\hat{J})$ is the radius of the circular orbit with angular momentum $\hat{J}=(\tilde{J}^2+J_{\rm min}^2)^{1/2}$, where $\tilde{J}=|J_\phi| + k_r J_r + k_z J_z$, and $k_r$ and $k_z$ are dimensionless coefficients. The reason we use $\hat{J}$ instead of $J_\phi$ as argument of $R_c$ is that the former gives a value that better represents the average radius of a star with given angular momentum $J_\phi$ (e.g.\ a star with $J_\phi \simeq 0$ will not in general stay close to $R=0$ if the other two actions are large). $J_{\rm min}$ is a parameter introduced to avoid a pathological behaviour of the DF in the case of a cuspy potential when epicyclic frequencies tend to infinity as $R\to0$. $\sigma_{\rm min}$ is a minimum value of velocity dispersion that is added in quadrature in Eqs.~\eqref{eq:sigmar} and \eqref{eq:sigmaz} in order to avoid the pathological situation when the velocity dispersions drop so rapidly with radius that the value of DF at $J_r = J_z = 0$ increases indefinitely at large $J_\phi$. We choose $k_r=1.0$, $k_z=0.25$ (in this way $\tilde{J}$ is approximately constant across an energy surface - see equation 16 and related discussion in \citealt{Vasiliev2019}), $J_{\rm min}= 10 \kms \kpc$ and $\sigma_{\rm min}=2\kms$. $\kappa(R_c)$ and $\nu(R_c)$ are the radial and vertical epicyclic frequencies at radius $R_c$, and $\Omega(R_c)$ is the angular frequency.

The NSD has a total of 5 free parameters that we fit to the data: $\{M_{\rm NSD}, R_{\rm disc}, H_{\rm disc}, \sigma_{r,0}, R_{\rm \sigma,r} \}$. $M_{\rm NSD}$ is the total mass of the NSD, which is specified through the overall normalisation to the surface density profile $\Sigma_0$. $R_{\rm disc}$ controls the radial scale-length of the density profile, while $\sigma_{r,0}$ and $R_{\sigma,r}$ control the central radial velocity dispersion and the radial scale of the (approximately exponential) radial velocity dispersion profile.\footnote{Here radial velocity refers to the radial velocity with respect to the Galactic Centre, not the line-of-sight velocity.} Thus, the density and velocity dispersion profiles can be varied independently. Increasing $\sigma_{r,0}$ while keeping fixed the other parameters makes the disc hotter (and therefore decreases the amount of rotation in the disc). Decreasing $R_{\sigma,r}$ while keeping the other parameters fixed makes the disc colder in the outer parts relative to the inner parts. $H_{\rm disc}$ sets the vertical scale-height and also controls the vertical velocity dispersion since the latter is determined by the self-gravity of the disc and hence by its scale-height. Increasing $H_{\rm disc}$ will make the disc vertically hotter and thicker. The azimuthal ($\phi$) velocity dispersion is uniquely linked to the radial one.

\subsection{Iterative procedure} \label{sec:iterative}

Given $\Phi_{\rm NSC}(R,z)$ and $f(\bfJ)$, the model is (in theory) completely specified by the requirement of self-consistency. But in practice it is not trivial to find the density/potential pair $\{\rho_{\rm NSD}(\bfx),\Phi_{\rm NSD}(\bfx)\}$ that is implied by this requirement. We find this pair using the iterative procedure introduced by \citet{Binney2014} implemented in the self-consistent galaxy modelling module of the software package \textsc{Agama} \citep{Vasiliev2019}.

Given a gravitational potential $\Phi(\bfx)=\Phi_{\rm NSC}+\Phi_{\rm NSD}$ and $f(\bfJ)$, \textsc{Agama} can calculate the DF $f(\bfJ(\bfx,\bfv))$ and the density $\rho_{\rm NSD}(\bfx)=\int \di^3 \bfv \, f(\bfJ(\bfx,\bfv))$ as a function of ordinary phase-space coordinates.\footnote{The standard method for estimating $\bfJ$ from $(\bfx,\bfv)$ is the `St\"ackel fudge' introduced by \citealt{Binney2012} with the refinements of \citealt{Vasiliev2019}; see also \citealt{Sanders2016} for a review of action estimation methods} A model is said to be self-consistent if the gravitational potential calculated from $\rho_{\rm NSD}(\bfx)$ via Poisson's equation coincides with the $\Phi_{\rm NSD}(\bfx)$ given at the beginning. The gravitational potential $\Phi_{\rm NSD}(\bfx)$ that accomplishes this is in general not known a priori, and it is determined through the iterative procedure.

The procedure works as follows. We start with an initial guess for the gravitational potential $\Phi_0(\bfx)=\Phi_{\rm NSC} + \Phi_{\rm NSD,0}$ and use this guess and $f(\bfJ)$ to evaluate $\rho_{\rm NSD,1}(\bfx)$. From this density and Poisson’s equation one recovers a new estimate of the NSD potential $\Phi_{\rm NSD,1}$, which is used to re-evaluate the densities and find an improved potential $\Phi_{\rm NSD,2}$. This sequence of densities and potentials usually converges after $\sim 5$ iterations \citep{Binney2014}. The model is then complete and ready to predict any observable. $\Phi_{\rm NSC}$ and $f(\bfJ)$ are kept fixed during the iterative procedure. As an initial guess we use the density/potential generated by the best-fitting NSD model (model 3) of \citet{Sormani2020a}. 
At each iteration, the density and the potential of the NSD are evaluated on a cylindrical grid which is logarithmically spaced in radius between $\texttt{RminCyl}=10^{-3} \kpc$ and $ \texttt{RmaxCyl}=10 \kpc$ with $\texttt{sizeRadialCyl}=50$ radial points, and vertically between $\texttt{zminCyl}=5 \times 10^{-3} \kpc$ and $\texttt{zmaxCyl}=1 \kpc$ with $\texttt{sizeVerticalCyl}=30$ points (see \textsc{Agama} documentation). The entire procedure requires $\sim 5\min$ on an 8-core laptop.

\subsection{Coordinate systems} \label{sec:coordinate}

It is well known (and unfortunate) that the origin of the Galactic Coordinate system $(l,b)=(0,0)$ does not coincide with the location of Sgr~A* which is believed to mark the ``true'' Galactic Centre, $(l_{\rm Sgr\,A*},b_{\rm Sgr\,A*})=(-0.05576432, -0.04616002)^\circ$. This offset was taken into account in the design of the KMOS NSD survey of \citet{Fritz2021}. In order to deal with this fact, we use two coordinate systems in this paper.

The first is a right-handed Cartesian Galactocentric coordinate system $(X,Y,Z)$ oriented such that the $XY$ plane is the Galactic plane and $Z$ points towards the North Galactic Pole. This system is at rest with respect to the Galactic Centre and is used as the basis to define the usual Galactic Coordinates, so the origin $(X,Y,Z)=0$ corresponds to $(l,b)=(0,0)$. The Sun is assumed to be located at $(X_{\rm \odot},Y_{\rm \odot}, Z_{\rm \odot})=(0.0,-8.2,0.025)\kpc$ and to have total velocity (Local Standard of Rest + peculiar) $(V_{x \rm \odot},V_{y \rm \odot}, V_{z \rm \odot})=(-249,10,7)\kms$ in this system%
\footnote{Note the swapping of $X$ and $Y$ axes with respect to other commonly used convention, e.g., from \textsc{Astropy}.}
\citep[e.g.][]{Bland-Hawthorn2016}. The position of Sgr~A* in this system is assumed to be $(X_{\rm Sgr\,A*},Y_{\rm Sgr\,A*},Z_{\rm Sgr\,A*})=(7.98,0.0,-6.6)\pc$, so that its Galactic Longitude and Latitude coincide with the observed ones.

The second is a Cartesian coordinate system $(x,y,z)$ which is centred on Sgr~A*. It is related to the previous coordinate system by a simple translation:
\begin{align}
 x & = X  - X_{\rm Sgr\,A*} \,,\nonumber \\
 y & = Y  - Y_{\rm Sgr\,A*} \,,\\
 z & = Z  - Z_{\rm Sgr\,A*} \,.\nonumber 
\end{align}
 We assume that both the NSD and NSC are centred on the origin of the $(x,y,z)$ system and that $z$ coincides with their axis of symmetry.

\section{Fitting procedure} \label{sec:fitting}

We compare the model and the data using the normalised line-of-sight velocity and proper motion distributions in each of the 24 KMOS fields displayed in Figure~\ref{fig:kmos_01}. The comparison is purely based on the kinematics and neglects any photometric information. 

We model the distributions as the sum of two contributions: (i) the NSD, and (ii) the contaminating background due to the Galactic Bar/Bulge and Disc (hereafter we refer to this simply as the ``Bar'' for simplicity). We fit the predicted distributions using a likelihood (Equation \ref{eq:logP}) that is a function of the 5 free parameters of the NSD model, $\theta = \{M_{\rm NSD}, R_{\rm disc}, H_{\rm disc}, \sigma_{r,0}, R_{\rm \sigma,r} \}$ (see Section \ref{sec:NSD}). The following subsections describe how we calculate the likelihood of the model.

\subsection{Definitions}

Let us denote the three kinematic observables with the notation $o_k$, where $k=\{1,2,3\}$ and $o_1 = \vlos$, $o_2=\mu_l$ and $o_3=\mu_b$. Consider a star $i$ in a KMOS field $j$. We call $p(o_{k,i} | j, k, \theta)$ the probability that the star has kinematic observable $o_{k,i}$ given that it is located in the field $j$, that it has a measurement of the observable $k$ and given the parameter set $\theta$. This probability satisfies 
\begin{equation}
    \int_{-\infty}^{+\infty} p(o_k | j,k,\theta)\, \di o_k = 1.
\end{equation} 
In our model, each star belongs to either the NSD or the Bar. Therefore we write:
\begin{align} \
    p(o_{k,i} | j,k, \theta) & = p(\NSD | j,k, \theta) \,  p(o_{k,i} | j, k,
    \theta, \NSD) \nonumber \\ 
    & + p(\BAR | j, k,\theta) \, p(o_{k,i} | j, k,\theta, \BAR) \,,\label{eq:p1}
\end{align}
where 
\begin{itemize}
    \item $p(\NSD | j,k, \theta)$ is the probability that the star belongs to the NSD given that it is in field $j$ and that it has a measurement of the observable $k$;
    \item $p(o_{k,i} | j,k,\theta,\NSD)$ is the probability that the star has kinematic observable $v_i$ given that it is located in the field $j$, that it has a measured observable $k$ and that it belongs to the NSD;
\end{itemize} 
and so on with obvious notation. These probabilities satisfy \begin{equation}
    \int_{-\infty}^{+\infty} p(o_k | j, k, \theta, \NSD)\, \di o_k = 1\,,
\end{equation} and \begin{equation} p(\NSD | j,k, \theta) + p(\BAR | j,k, \theta) = 1\,. \label{eq:P2}\end{equation} 

\subsection{Calculation of  $p(o_k | j,k,\theta, \NSD)$} 
\label{sec:nsdvdf}

We calculate $p(o_k | j,k,\theta, \NSD)$ using the self-consistent NSD model described in Section \ref{sec:scm}. The procedure is as follows:
\begin{enumerate}
    \item Generate $2 \times 10^7$ stellar samples from the model using the \textsc{Agama} built-in sampling tool. This is done only the first time (i.e.\ for the first evaluation of the likelihood), and for subsequent models we use the same fixed set of samples reweighted by the new model to avoid small random fluctuations between models.
    \item Calculate the $(l,b,o_k)$ position and the distance $d$ of every sampled star.
    \item Retain only the stars that in the $(l,b)$ plane fall within a radius of $0.07\degree$ from the centre of the field $j$.
    \item  Construct a 1D Kernel Density Estimation (KDE) of the $o_k$ distribution of the retained stars, weighted by their mass and by the selection function $S_{k,j}(d)$ (Section~\ref{sec:selection}). The KDE is constructed using a Gaussian kernel, and the bandwidth $b$ is estimated using the \citet{Scott1992} rule\footnote{The Scott's rule says that $b=n_{\rm eff}^{-1/(D+4)}$ where $n_{\rm eff}=(\sum_i w_i)^2/\sum_i w_i^2$ is the effective number of datapoints, $w_i$ are their weights and $D$ is the number of dimensions.}. We use the KDE implementation from the {\sc KDEpy} package.\footnote{\url{https://kdepy.readthedocs.io/en/latest/index.html}} The probability distribution estimated by the KDE constitutes $p(o_k | j, k,\theta, \NSD)$.
\end{enumerate}

\subsection{Calculation of $p(o_k | j,k, \theta, \BAR)$} \label{sec:background}

The background due to the Galactic Bar/Bulge and the Disc is explicitly taken into account using an $N$-body model from \citet{Portail2017} [hereafter P17].  These authors constructed dynamical models of the Milky Way Bar by integrating an $N$-body system and slowly adjusting the masses of the particles until the time-averaged density field and other model observables converged to prescribed data, using the made-to-measure method \citep{Syer1996,deLorenzi2007}. The P17 models are constrained to reproduce a variety of stellar density and kinematic data, and they build upon previous reconstructions of the 3D Bar density from red clump giant star counts \citep{Wegg2013,Wegg2015}. P17's overall best-fitting model had a pattern speed of $\Omega_{\rm p} = 40\kms\kpc^{-1}$. The pattern speed is one of the most important parameters of the Bar since it sets the location of the resonances. More recently, there has been evidence for somewhat lower values of $\Omega_{\rm p}$ \citep{Clarke2019,Binney2020a,Chiba2021b,Clarke2021}.  Here we consider the P17 model with $\Omega_{\rm p} = 37.5\kms\kpc^{-1}$ which was found to be a good match to the VIRAC proper motions in \citet{Clarke2019} and, with gas dynamical modelling, to the observed distribution of interstellar gas in the $(l,\vlos)$-diagram  \citep{Sormani2015a,Li2016,Li2021}. We neglect spiral arms in the Galactic Disc, which contain too few stars to have a significant impact on the normalised kinematic histograms used in our fitting procedure \citep{Nogueras-Lara2021d}. 

\begin{figure*}
	\includegraphics[width=\textwidth]{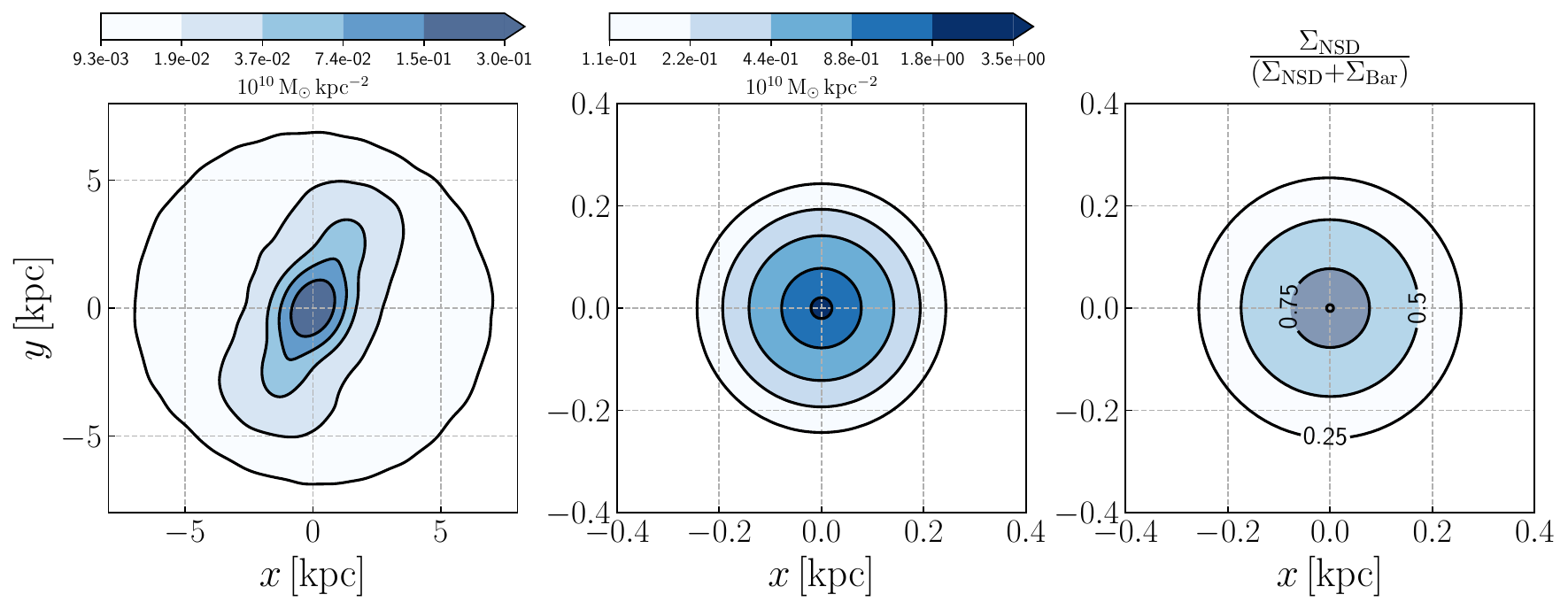}
    \caption{Top-down surface density of the \citet{Portail2017} model of the Galactic Bar (left), of the fiducial model of the NSD (middle), and of the ratio between the two (right). The ratio illustrates how prominent the Milky Way's NSD would be if we were to see it in an external galaxy. Note that the $(x,y)$ scale in the middle and right panels is much smaller than in the left panel. The Sun is at $(x,y) = (0,-8.2\kpc)$. The Galactic Bar model also includes the Galactic Disc (see text in Section \ref{sec:background}). Note that the ratio in the right panel is almost perfectly round, because the Bar is nearly axisymmetric on the scale of the NSD.}
    \label{fig:bar}
\end{figure*}

\begin{figure}
	\includegraphics[width=\columnwidth]{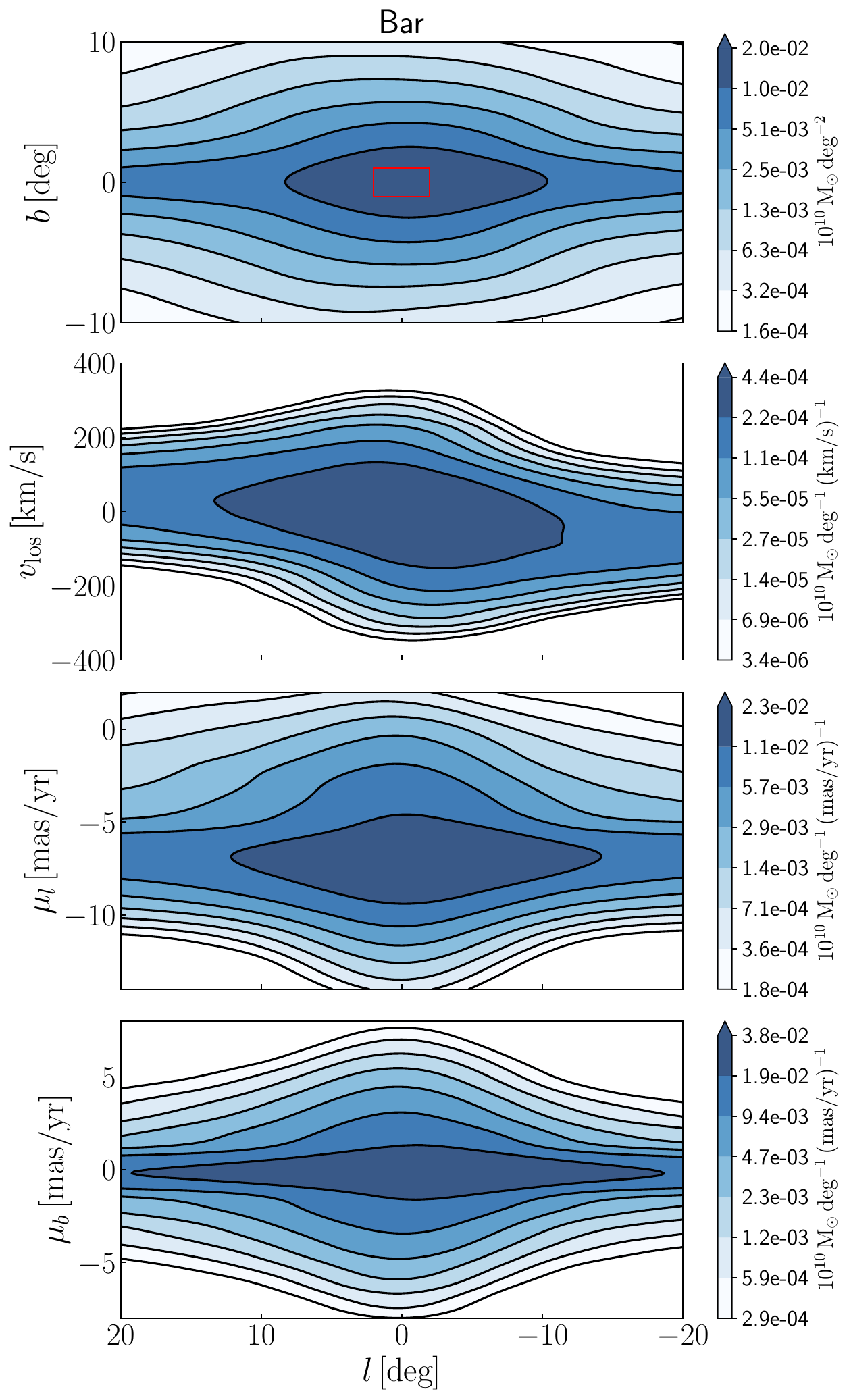}
    \caption{The contaminating background due to the Galactic Bar, calculated using the $N$-body model of \citet{Portail2017}, projected to various observational spaces. Contours are geometrically spaced every factor of $2$. The red square in the upper panel indicates the region $|l|<2\degree$, $|b|<1\degree$ where the NSD is located.}
    \label{fig:background}
\end{figure}

The P17 model is used here without the NSD-like central mass concentration (see their Sections 7.3 and 10.2), and so it is well-suited to complement our NSD model without creating issues of double counting. Figure~\ref{fig:bar} shows the face-on surface density of the Bar model adopted here, and Figure~\ref{fig:background} shows how it appears in various observational spaces.

In order to construct $p(o_k | j, k,\theta, \BAR)$ we proceed as follows:
\begin{enumerate}
    \item Calculate the $(l,b,o_k)$ position and the distance $d$ of every $N$-body stellar particle in the P17 model. The model has a total of $\simeq 750,000$ stellar particles within the solar circle ($R<8.2\kpc$).
    \item Construct a 3D KDE of the distribution of particles in $(l,b,o_k)$ space, weighted according to their mass and according to the selection fraction $S_{k,j}(d)$. Again, we use Gaussian KDEs and estimate the bandwidth using Scott's rule.
    \item $p(o_k| j, k,\theta, \BAR)$ is obtained by evaluating the KDE as a function of $o_k$ with $(l,b)$ fixed at the location of the centre of the field $j$.
\end{enumerate}
The procedure is repeated for every field $j$ since the selection fraction $S_{k,j}$ is field-dependent.

\subsection{Calculation of $p(\NSD | j, k, \theta)$ and $p(\BAR | j, k, \theta)$}
\label{sec:fnsd}

We assume that:
\begin{align} \label{eq:ratio}
   p(\NSD | j, k, \theta) = \frac{\tilde{\Sigma}_{\NSD,j,k}}{\tilde{\Sigma}_{\NSD,j,k} + \tilde{\Sigma}_{\BAR,j,k}} \,,
\end{align}
where
\begin{align}
    \tilde{\Sigma}_{\NSD,j,k} & = \int_0^\infty S_{k,j}(s) \rho_{\NSD}(s) \di s \label{eq:sigma1} \\
    \tilde{\Sigma}_{\BAR,j,k} & = \int_0^\infty S_{k,j}(s) \rho_{\BAR}(s) \di s \label{eq:sigma2}
\end{align}
are the surface densities on the plane of the sky weighted by the selection fraction. Here, $s$ is the distance along the line-of-sight centred on field $j$ and $\rho_{\rm NSD}$ and $\rho_{\rm BAR}$ are the volume density of the NSD and BAR respectively. Note that Equations~\eqref{eq:sigma1} and \eqref{eq:sigma2} reduce to the normal formulas for the surface density when $S_{k,j}(d)\equiv 1$.

The integrals in \eqref{eq:sigma1} and \eqref{eq:sigma2} are evaluated as follows. For the NSD, we proceed as in Section~\ref{sec:nsdvdf} until step (iv), and then we sum the mass of all particles in the field $j$ weighting by the selection fraction. For the Bar the procedure is slightly different because of the much lower number of particles: we proceed as in Section~\ref{sec:background} until step (ii), then we construct a 2D KDE of the distribution of particles in $(l,b)$ space, weighted by their mass and by the selection fraction. We then evaluate this KDE at the centre of field $j$ to obtain $\tilde{\Sigma}_{\BAR,j,k}$. Once $p(\NSD | j, k, \theta)$ is known, $p(\BAR | j, k, \theta)$ is obtained from relation~\eqref{eq:P2}.

\subsection{Prior on the model parameters} \label{sec:prior}

We assume no prior (i.e., uniform prior on the log) on the four parameters $\{M_{\rm NSD}, R_{\rm disc}, H_{\rm disc}, \sigma_{r,0}\}$, 
while we assume a broad Gaussian prior on $\log(R_{\rm \sigma,r})$:
\begin{equation}
    \mathcal{P}(\theta) = \frac{1}{2\sqrt{2\pi}} \exp \left[ -\frac{1}{2} \left( \frac{\log(R_{\rm \sigma,r}/1 \kpc)}{2}  \right)^2\right]\,,
\end{equation}
This is done to prevent this parameter from becoming unrealistically large in our fitting procedure (as we will see in Section~\ref{sec:around}, we are only able to obtain a lower limit but no upper limit on this parameter).

\subsection{Taking into account observational errors} \label{sec:err}

The observational errors on the line-of-sight velocities in our sample are typically negligible, but those on proper motions are not (see bottom row in Figure~\ref{fig:kmos_02}). To take into account observational uncertainties, we use the following simple Monte Carlo procedure. For each star in our sample we make $N_{\rm err}=100$ copies, each having $\vlos$ drawn from a 1D Gaussian distribution with the measured $\vlos$ as mean and with $v_{\rm los, err}$ as standard deviation. We repeat this for the components of proper motion. We then fit the data by calculating the likelihood on the augmented sample of stars that contains $N_{\rm err}$ times more stars than the original sample (see next section). Although this simple approach of taking into account the errors might sometimes lead to slight biases when the errors are large \citep[e.g.][]{Fritz2018}, the quality cut on the proper motions used in this paper should avoid this issue.

\subsection{Likelihood function} \label{sec:likelihood}
We calculate the total likelihood as 
 \begin{align} \label{eq:logP}
      \log P & = \log \mathcal{P}(\theta) + \sum_{k=1}^3\sum_{j=1}^{24} \sum_{i=1}^{N_{k,j}} \log \tilde{p}(o_{k,i} | j,k,\theta) \,,
 \end{align}
 where 
 \begin{enumerate}
 \item the sum over $k$ is a sum over the three observables $o_1 = \vlos$, $o_2=\mu_l$ and $o_3=\mu_b$;
 \item the sum over $j$ represents the sum over the 24 fields;
 \item the sum over $i$ runs over all stars in the sample within each field $j$ that have the observable $o_k$ defined. The numbers $N_{k,j}$ depend on $j$ because the number of stars is in general different for each field, and on $k$ because the number of stars with a given observable within each field is in general different for each observable (e.g.\ a star might have $\vlos$ but not $\mu_l$, or vice versa);
 \item $\mathcal{P}(\theta)$ is the prior on the model parameters (see Section~\ref{sec:prior});
 \item we have defined
 \begin{equation} \tilde{p}(o_{k,i} | j,k,\theta) = \frac{1}{N_{\rm err}} \sum_{n=1}^{N_{\rm err}} p(o_{k,i,n} | j,k,\theta)\,,
 \end{equation}
 where $o_{k,i,n}$ is the kinematic observable $o_k$ for the $n$-th copy of the star $i$ generated using the Monte Carlo procedure described in Section~\ref{sec:err}. This is a simple way to take into account observational errors.
 \end{enumerate} 

The approach behind Equation \eqref{eq:logP} neglects correlations between $\vlos$ and $\mu_l$, i.e.\ we assume that the probability of a star having a certain $\vlos$ is independent of its $\mu_l$. Although in principle it would be better to use the joint likelihood $p(v_{{\rm los},i},\mu_{l,i},\mu_{b,i}|j,k,\theta)$, we have chosen to avoid this to simplify the approach and avoid unnecessary computational complications. We have checked a posteriori that the models give an adequate representation of the data in the $\vlos\mhyphen\mu_l$ plane.

\section{Results} \label{sec:results}

\subsection{Fiducial model} \label{sec:fiducial}

Table \ref{tab:1} lists the parameters of our fiducial model. This model is obtained by maximising the likelihood~\eqref{eq:logP} using a standard Powell algorithm.

\begin{figure}
	\includegraphics[width=\columnwidth]{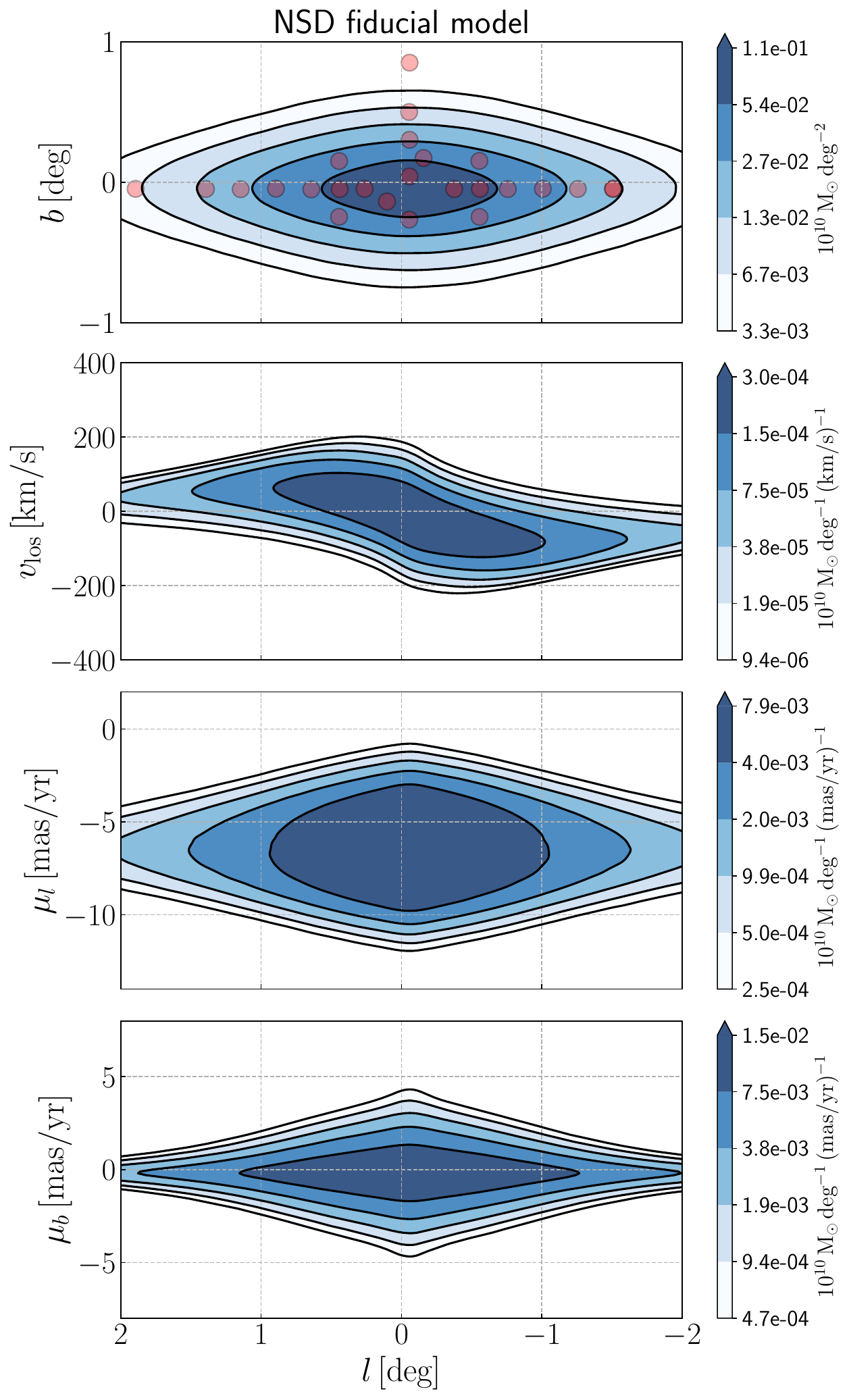}
    \caption{The fiducial NSD model projected to various observational spaces. The red circles in the top panels show the KMOS fields. Contours are geometrically spaced every factor of $2$. Note that the Galactic Bar background varies on scales much larger than the NSD (compare with Figure~\ref{fig:background}). These figures do not take into account the selection functions, they are obtained by simply binning the model particles in the various planes. One degree corresponds to roughly $140\pc$ at the distance of the Galactic Centre. The NSC is not included in this figure.}
    \label{fig:nsd1}
\end{figure}

Figure \ref{fig:nsd1} shows the fiducial model projected to various observational spaces. Figures~\ref{fig:scm_hist_0}-\ref{fig:scm_hist_3} compare in detail the fiducial model to the data in each KMOS field. The normalised kinematic distributions shown in these plots are the basis of our fitting procedure.\footnote{Note that the histograms in these figures are only shown in these figures to facilitate comparison since the fitting is done on a star-by-star basis and does not require binning, see Section~\ref{sec:fitting}.} There is generally good agreement between the model and the data. The distributions are typically made of a narrower component due to the NSD and a broader distribution due to the background of the Galactic Bar. The Bar dominates the tails of both the line-of-sight velocity and proper motion distributions in all fields, including the central ones. The agreement is surprisingly good if we consider that we used the P17 Bar model, which was fitted to a different dataset, without rescaling or adapting it in any way.

The $\mu_l$ distributions of the NSD in the central fields show a characteristic double-peaked shape \citep{Trippe2008,Schodel2009,Chatzopoulos2015b,Shahzamanian2021}. This is due to rotation and can be understood by considering the limiting case of a cold disc of stars in purely circular motion: at $l=0$ stars move perpendicularly to the line-of-sight, and stars rotating in one sense (and placed in the front side of the NSD) will give rise to one peak, and stars rotating in the opposite sense (and placed on the back side of the NSD) to the other peak. The peaks are broad because our fiducial NSD model is rather hot (see $\sigma_{r,0}$ in Table~\ref{tab:1}). The peaks are stronger at $l=0$ and blend together as we move to higher longitudes due to geometric effects (the velocity of a particle in circular orbit is exactly perpendicular to the line of sight only at $l=0$, but is not as we move away from it). When looking at the total distributions (NSD+Bar), the peaks in many fields tend to be washed out due to the significant background of the Bar. The peaks become more (less) pronounced for models with smaller (larger) dispersion $\sigma_{r,0}$ because the NSD becomes colder (hotter).

The $\mu_l$ distributions of the Galactic Bar are in general asymmetric and skewed towards lower (more negative) proper motions, with a shoulder in the range $\mu_l=-5$ to $0 \masyr$. This is mostly due to a purely geometric effect caused by the fact that proper motions are defined as the tangential velocity divided by the distance. This effect would be present even if the Bar were axisymmetric and if the selection fraction were unity ($S_{k,j}=1$). It can be understood by considering the limiting case of a disc of stars in purely circular motion with a given velocity curve: if we plot $\mu_l$ as a function of distance along the line-of-sight, we find a curve that is not symmetric around the Galactic Centre, and this gives rise to the asymmetry. We give a brief illustration of this effect using a toy model in Appendix~\ref{sec:appendix_skewed}.

The normalised kinematic distributions seem to be well reproduced under the assumption of axisymmetry and betray no obvious asymmetric residuals that would suggest the presence of a nuclear bar. We have checked that this is not due to the selection function (which could potentially conceal asymmetries in the NSD by assigning them to non-axisymmetric properties of the Bar population): a model obtained by maximising the likelihood while completely ignoring the selection function (i.e.\ assuming that the selection function is constant, $S_{k,j} \equiv 1$, see Section \ref{sec:around}) still reproduces the normalised kinematic distributions well and shows no obvious systematic residuals. Note that an axisymmetric NSD would not be in conflict with the observed gas asymmetry in the Central Molecular Zone, because the latter is likely caused by processes that mostly do not affect the NSD such as hydrodynamic instabilities and stellar feedback \citep{Sormani2018c}. However, the signal of a nuclear bar could be rather weak - experiments we have conducted by scaling down an $N$-body large-scale bar suggest that a signature of the presence of a nuclear bar should be a slight longitudinal asymmetry in the peak of the $\mu_l$ distributions. Determining whether the NSD is truly axisymmetric will probably require a larger statistical sample of data as well as more detailed theoretical investigations of what the signature of a nuclear bar would be.

Figure~\ref{fig:1D} shows radial density profiles in the plane $z=0$. The fiducial model of the NSD dominates the density at $R<300\pc$, which a-posteriori justifies our choice of not including the gravitational potential of the Bar in our self-consistent modelling and only treat the Bar as a background contaminant. The NSC dominates only in the innermost few parsecs, as expected.

\begin{figure}
	\includegraphics[width=\columnwidth]{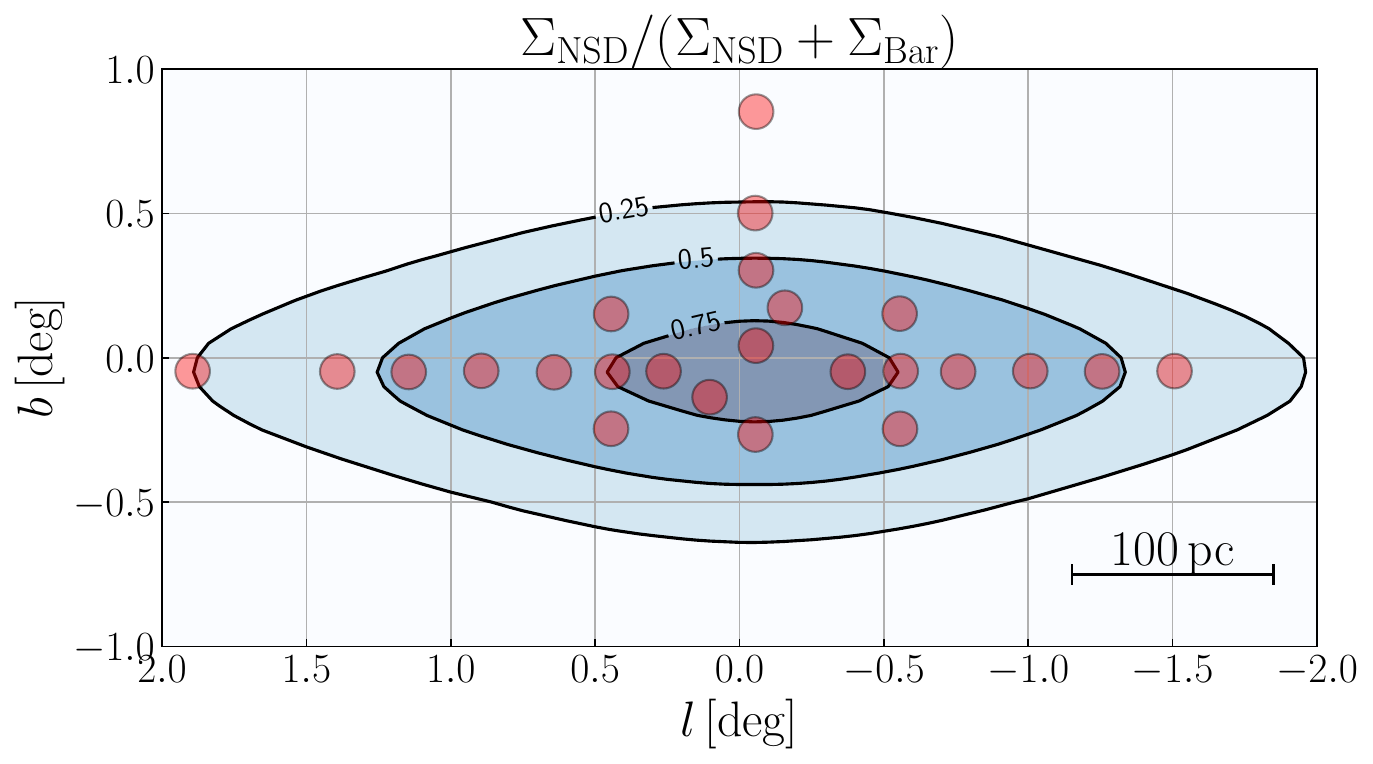}
    \caption{Ratio between the surface density (on the plane of the sky) of the fiducial NSD model and the Galactic Bar. This ratio gives an indication of the fraction of stars that belong to the NSD at each position in the sky. The values calculated by taking into account the full selection function are not very different (see Table \ref{tab:2}). Contours show $25\%$, $50\%$ and $75\%$ levels. Contamination from the Bar is significant over most fields.}
    \label{fig:ratio}
\end{figure}

\begin{figure}
	\includegraphics[width=\columnwidth]{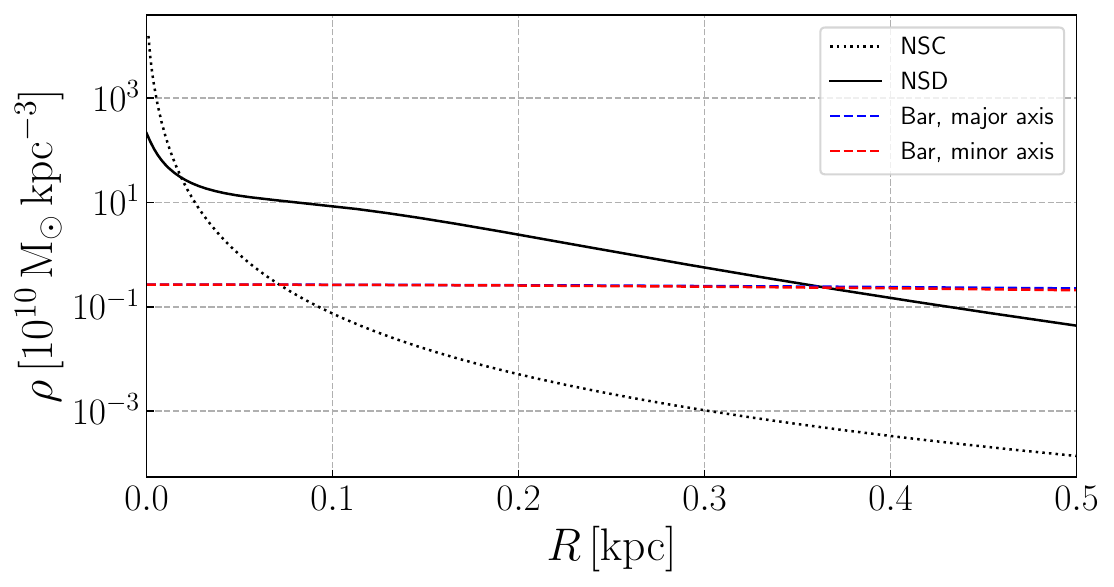}
    \caption{Radial density profiles in the $z=0$ plane. \emph{Dotted line:} the NSC. \emph{Full line:} the NSD fiducial model. \emph{Dashed lines}: the Galactic Bar model of \citet{Portail2017} along the Bar's minor and major axis. There is only one line for the NSC and NSD because these two components are axisymmetric, and two lines for the Bar since this component is not axisymmetric, although they are almost identical in this radial range. The fiducial NSD model dominates the density at $R<300\pc$.}
    \label{fig:1D}
\end{figure}

Contamination due to the Bar is substantial in most fields. The second column in Table~\ref{tab:2} gives the probability that a star in our sample belongs to the NSD according to our fiducial model. In the central fields this is around $70\mhyphen80\%$, and drops as we move away from the centre. The third column gives the probability that a star in our sample belongs to the NSD for the subset of stars with $H-K>1.3$. According to our model this probability is slightly higher but not very different from that for the whole sample. This suggests that there is probably a significant number of Bar/Bulge stars moving through the NSD. The fourth column gives the ratio between the surface density of the NSD and the total surface density of Bar+NSD. Although this simple ratio does not take into account the selection function from Section~\ref{sec:selection}, we can see that it gives values very similar to the second column which does take the selection function into account, and is therefore a very good proxy of the actual probability. This ratio is displayed in Figure~\ref{fig:ratio}, which can therefore be considered a good indication of the contamination due to the Bar as a function of position in the sky.

\begin{table}
\caption{Parameters of our fiducial model. This maximises the likelihood given by Eq.~\eqref{eq:logP}.}
\centering
\begin{tabular}{ccccc}  
\toprule
		 $M_{\rm NSD}$ &  $R_{\rm disc}$ &  $H_{\rm disc}$ &  $\sigma_{r,0}$ &  $R_{\rm \sigma,r}$ \\
		 $[10^{8}\Msun]$ &  $[\rm pc]$ &   $[\rm pc]$ & $[\kms]$ & $[\rm pc]$ \\
\midrule
 9.7  & 74  & 26  & 75 & 100 \\
\bottomrule
\end{tabular}
\label{tab:1}
\end{table}

\subsection{Parameter search} \label{sec:around}

\begin{figure*}
	\includegraphics[width=1.0\textwidth]{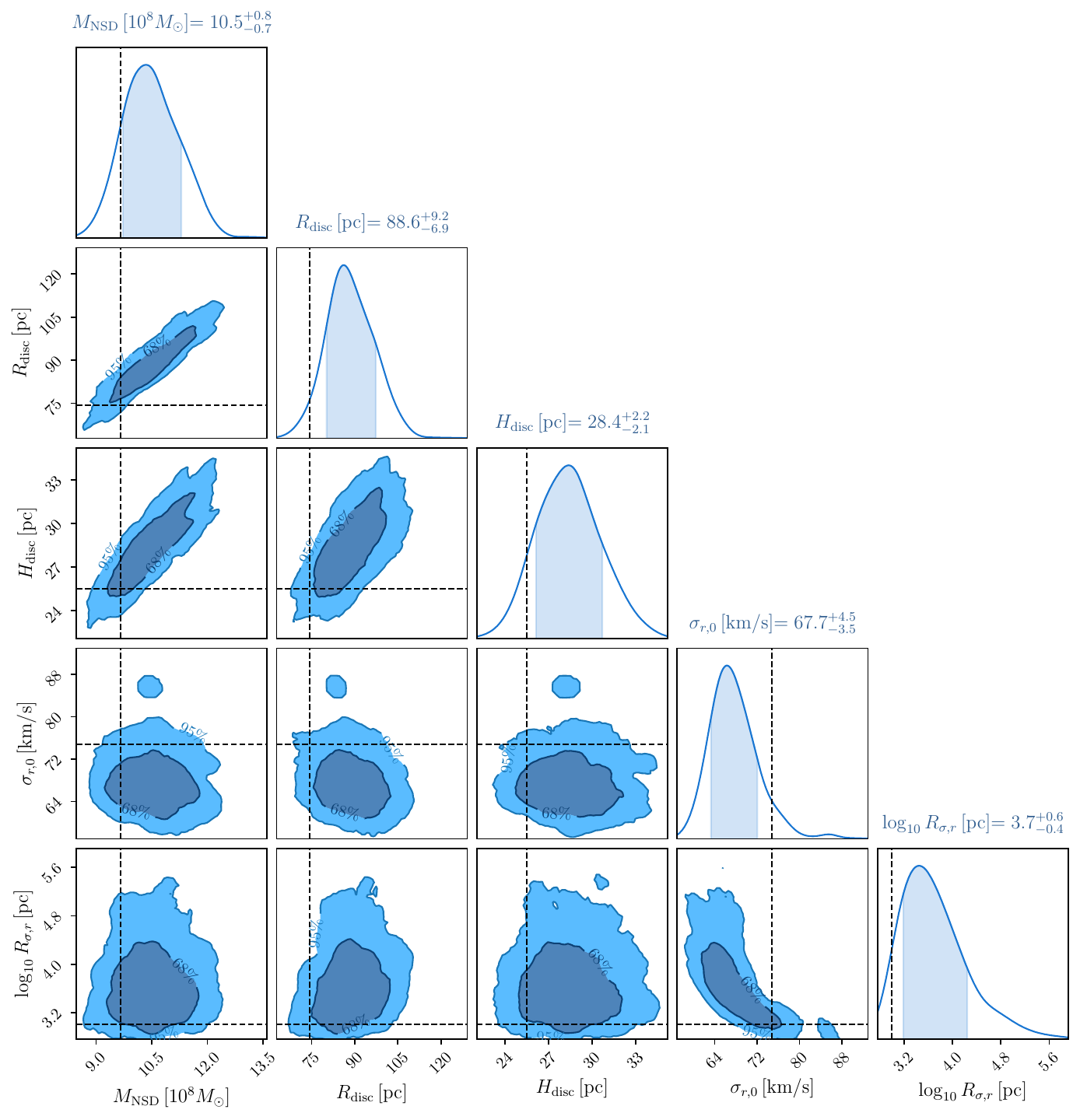}
    \caption{Posterior distributions obtained by running an MCMC using the likelihood in Equation~\eqref{eq:logP}. The contours contain 68 and 95 per cent of the samples of the chain. The black dashed lines show the parameters of the fiducial model (Table~\ref{tab:1}). The prior on the parameters are discussed in Section~\ref{sec:prior}.}
    \label{fig:mcmc}
\end{figure*}

To explore the parameter space we run a Monte-Carlo Markov Chain (MCMC) using the likelihood \eqref{eq:logP}. We use the package {\sc emcee} \citep{Foreman-Mackey2013} and run the chain with 64 walkers and 240 steps. The results are shown in Figure~\ref{fig:mcmc}.

The marginalised 1D distributions for each parameter are reported in Figure~\ref{fig:mcmc}. The total mass ($M_{\rm NSD}$), radial scale-length ($R_{\rm disc}$), vertical scale-height ($H_{\rm disc}$) and central velocity dispersion ($\sigma_{r,0}$) are all well constrained by the data. The parameter $R_{\sigma,r}$, the scale-length of the radial velocity dispersion profile, is less constrained. We only obtain a lower limit on this parameter, i.e. it cannot be too small. This is because if the velocity dispersion drops too rapidly with radius, the velocity dispersion of the model is too small compared to the data in the outer parts of the NSD.

Figure~\ref{fig:mcmc} also shows that there are some correlations between the parameters. The $M_{\rm NSD}$ vs.\ $R_{\rm disc}$ correlation means that in practice what is constrained is the surface density $\sim M_{\rm NSD}/R_{\rm disc}^2$. Similarly, the $M_{\rm NSD}$ vs.\ $H_{\rm disc}$ correlation implies that the edge-on surface density $\sim M_{\rm NSD}/(R_{\rm disc} H_{\rm disc})$ is being constrained. In both cases, by increasing the mass while decreasing the scale-length it is possible to keep the NSD contribution to the normalised histograms roughly at the same level compared to the Bar background, hence the correlation. The $R_{\rm disc}$ vs.\ $H_{\rm disc}$ correlation follows from the previous two. The $\sigma_{r,0}$ vs.\ $R_{\sigma,r}$ correlation likely arises because the velocity dispersion of the NSD is constrained to be fixed at around $\simeq 65 \kms$ in the outer fields, and one can keep this roughly constant by increasing (decreasing) the central $\sigma_{r,0}$ while simultaneously decreasing (increasing) $R_{\sigma,r}$.

The fiducial model described in Section~\ref{sec:fiducial} is not at the centre of the posterior distributions in Figure~\ref{fig:mcmc}. It turns out that there is a ridge in the 5D parameters space of models that all have essentially the same likelihood, and that fit the data equally well. This ridge also passes through the centre of the distributions in Figure~\ref{fig:mcmc}. We prefer the fiducial model presented in Section~\ref{sec:fiducial} rather than the model that goes through the centre of the distribution because, by eye, we find that it provides a slightly better match to the density distribution in previous photometric studies. In any case, the fiducial model should be taken as a representative model, but there is a range of almost equally plausible models.

\begin{figure*}
	\includegraphics[width=1.0\textwidth]{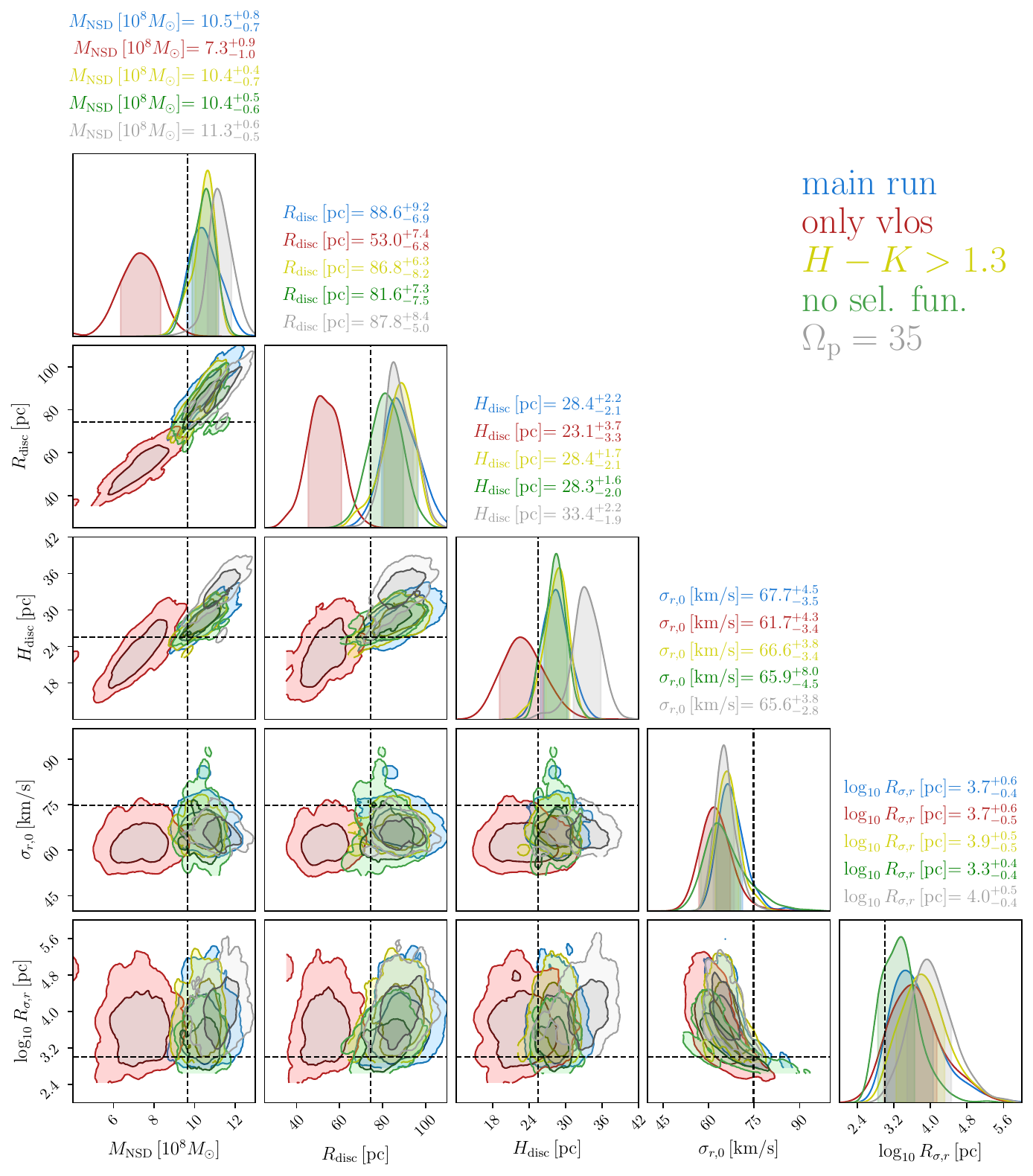}
    \caption{Posterior distributions obtained by running an MCMC under a variety of conditions. ``main run'' is the same as in Fig.~\ref{fig:mcmc}. ``$H-K>1.3$'' means that the models are fitted only using stars that are redder than this cut, and correspondingly using $(H-K)_{\rm cut}=1.3$ for all fields in step (ii) of the construction of the selection fraction, see Section~\ref{sec:selection}. ``only vlos'' are models fitted considering only the line-of-sight velocity of the stars and ignoring the proper motions. ``no selection function'' are models fitted with a uniform selection fraction, $S_{k,j}\equiv1$. 
    ``$\Omega_{\rm p}=35$'' are models fitted using a model of the Galactic Bar that has a slightly lower pattern speed than the main Bar model used in this paper. 
    }
    \label{fig:mcmc2}
\end{figure*}

\begin{figure*}
	\includegraphics[width=\textwidth]{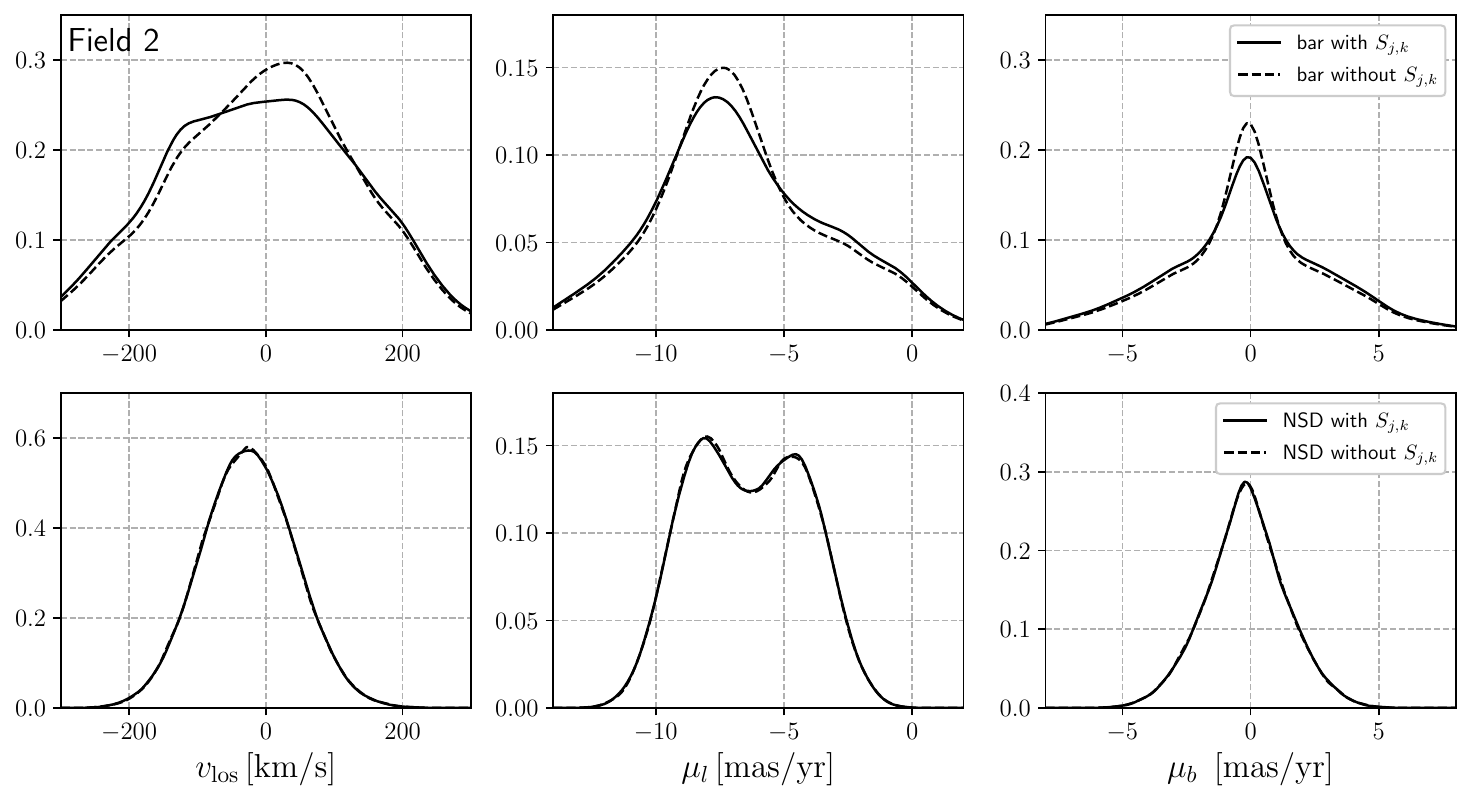}
    \caption{Impact of the selection function on our modelling procedure. Shown are the normalised kinematic histograms for the Bar (top panels) and the NSD fiducial model (bottom panels). Full lines are calculated taking into account the selection fraction $S_{k,j}$ (see Section~\ref{sec:selection}), while dashed lines are without taking it into account (i.e., with $S_{k,j}\equiv1$ identically). The effect is significant but not dramatic for the Bar, while it is negligible for the NSD.}
    \label{fig:kde}
\end{figure*}

Figure~\ref{fig:mcmc} only reports the statistical uncertainties. To get an estimate of the systematic uncertainties, we repeat the MCMC under different conditions. Figure~\ref{fig:mcmc2} reports the results of these experiments. We see that ignoring the selection function (i.e.\ assuming that we can see all stars along the line of sight with equal probability, green in Fig.~\ref{fig:mcmc2}) makes almost no difference to our results. Indeed, the impact of the selection function on the normalised kinematic histograms that form the basis of our fitting procedure is moderate (see Figure~\ref{fig:kde}). This suggests that the systematic uncertainty related to the selection criteria of our survey is small. Considering only stars that have $H-K>1.3$ (yellow in Fig.~\ref{fig:mcmc2}), which introduces a more stringent distance cut that excludes nearby stars in the Galactic Disc, also makes essentially no difference to the results. Surprisingly, we find that fitting the models using only the line-of-sight velocities while ignoring the proper motions (red in Fig.~\ref{fig:mcmc2}) gives posterior distributions for $M_{\rm disc}$ and $R_{\rm disc}$ that are inconsistent with those in the main run. Closer inspection reveals that there is only one parameter that is really inconsistent between the two runs, because the $M_{\rm disc}$-$R_{\rm disc}$ panel in the top-left of Figure~\ref{fig:mcmc2} shows that the ``only vlos'' 2D distribution follows the correlation discussed above which implies that $M_{\rm disc}/R_{\rm disc}^2\sim$ constant. This means that considering only the line-of-sight velocities favours an NSD with a smaller radius and approximately the same surface density (and therefore also a smaller total mass). The likely reason at the origin of this behaviour is that in the fields at high longitude ($|l|\gtrsim1^\circ$) the $\vlos$ distributions are reasonably well fitted by the Bar alone, so the fitting procedure prefers to eliminate the NSD contribution at large radius. When the information from the proper motion is added, the likelihood recognises that the Bar alone is not sufficient anymore to reproduce the distributions in the outer fields, and prefers to add a small contribution from the NSD at large radius to improve the fit.

Finally, we estimate the systematic errors associated with the adopted bar model. The ``$\Omega_{\rm p}=35$'' run repeats the fitting procedure with a different bar model from P17 that fits well the gVIRAC (VIRAC+Gaia DR2) proper motion data discussed in \cite{Clarke2021}. The main differences between this model and the main model used in the rest of this paper are that the former has a slightly lower pattern speed ($\Omega_{\rm p}=35 \kms\kpc^{-1}$ vs $\Omega_{\rm p}=37.5 \kms\kpc^{-1}$) and that it was originally fitted to the data using an NSD-like component (which we have removed for the purposes of this paper, see Sections 7.3 and 10.2 of P17) with a lower total mass ($1\times 10^9\, \Msun$ vs $2\times 10^9\, \Msun$), which affects the inner structure of the bar model. Thus, this different bar model tests uncertainties related to the mass distribution of the inner bulge as well as those related to the pattern speed of the bar. As can be seen from Figure~\ref{fig:mcmc2}, the results for most of the best-fitting NSD parameters are nearly unchanged when we use the different bar model. The main difference is that the latter favours a slightly larger scale-height ($H_{\rm disc}=33.4\pc$ vs $H_{\rm disc}=28.4\pc$ of the main run). This happens because the newer bar model has less pronounced wings in the $\mu_b$ distributions of the bar particles, which need to be partially compensated by a larger NSD scale-height. This feature of the newer bar model is probably related to the lower mass of the original NSD-like component rather than to its lower pattern speed. The differences in the NSD parameters obtained with the two bar models are treated as estimates of systematic uncertainties and are added in quadrature to the errors quoted in the abstract, discussion and conclusions.

\begin{table}
\caption{The columns in this table are defined as follows. \emph{Field} is the field number (see Fig.~\ref{fig:kmos_01}). $P(\NSD)$ is the probability that a star that has $\vlos$ in the survey of \citet{Fritz2021} belongs to the NSD (i.e., it is $P(\NSD|j,k,\theta)$ from Eq.~\ref{eq:ratio} where $k=1$ and $\theta$ are the parameters of the fiducial model). $P(\NSD)_{(H-K)>1.3}$ is the probability that a star that has $\vlos$ and $(H-K)>1.3$ in the survey of \citet{Fritz2021} belongs to the NSD (obtained by using Eq.~\ref{eq:ratio} with a modified $S_{k,j}$ that assumes $(H-K)_{\rm cut}=1.3$ in step ii in Section~\ref{sec:selection}). This quantity is not defined for field 5 since there are no stars with $(H-K)>1.3$ in this field. $\Sigma_{\NSD}/(\Sigma_{\rm Bar}+\Sigma_{\NSD})$ is the ratio between the surface densities in the plane of the sky of the fiducial NSD model and of the Bar (i.e., Eq.~\ref{eq:ratio} with $S_{k,j} \equiv 1$, what is displayed in Fig.~\ref{fig:ratio}). This ratio is a very good proxy of the actual probability.}.
\centering
\begin{tabular}{lccc}  
\toprule
			 Field & $P(\NSD)$ & $P(\NSD)_{(H-K)>1.3}$ & $\Sigma_{\NSD}/(\Sigma_{\BAR}+\Sigma_{\NSD})$ \\
\midrule
1 & 0.83 & 0.83 & 0.81 \\
2 & 0.68 & 0.69 & 0.65 \\
3 & 0.52 & 0.53 & 0.49 \\
4 & 0.26 & 0.28 & 0.24 \\
5 & 0.05 & N/D & 0.05 \\
7 & 0.78 & 0.79 & 0.76 \\
10 & 0.76 & 0.78 & 0.74 \\
12 & 0.70 & 0.71 & 0.67 \\
13 & 0.74 & 0.74 & 0.71 \\
14 & 0.71 & 0.73 & 0.66 \\
15 & 0.63 & 0.64 & 0.58 \\
16 & 0.57 & 0.19 & 0.49 \\
17 & 0.43 & 0.46 & 0.39 \\
18 & 0.76 & 0.77 & 0.74 \\
19 & 0.73 & 0.74 & 0.70 \\
20 & 0.69 & 0.69 & 0.66 \\
21 & 0.61 & 0.63 & 0.58 \\
22 & 0.53 & 0.55 & 0.48 \\
23 & 0.42 & 0.44 & 0.38 \\
24 & 0.64 & 0.65 & 0.61 \\
25 & 0.62 & 0.63 & 0.59 \\
26 & 0.64 & 0.65 & 0.61 \\
27 & 0.63 & 0.65 & 0.60 \\
28 & 0.24 & 0.27 & 0.22 \\
\bottomrule
\end{tabular}
\label{tab:2}
\end{table}

\section{Discussion} \label{sec:discussion}

\subsection{The mass of the Nuclear Stellar Disc} \label{sec:mass}

The total mass of the NSD according to the marginalised posterior distributions in Figure~\ref{fig:mcmc} is $M_{\rm NSD} = 10.5^{+1.1}_{-1.0} \times10^8 \,\Msun$. This is consistent with the previous determination from photometry by \citet{Launhardt2002}, who found $14.2 \pm 6 \times 10^8 \,\Msun$. We note that the two determinations are completely independent since ours is purely based on the normalised kinematics. Our determination is slightly higher than that of the Jeans modelling of \citet{Sormani2020a}, who found $6.9 \pm 2 \times 10^8 \,\Msun$, probably because our model here has a larger radius (see Fig.~\ref{fig:compare}). Finally, our mass is consistent with that required in gas dynamical models to reproduce the observed size of the CMZ gaseous ring (e.g.\ \citealt{Li2021} and references therein).

\subsection{Radial and vertical scale-heights} \label{sec:scaleheight}

\begin{figure*}
	\includegraphics[width=\textwidth]{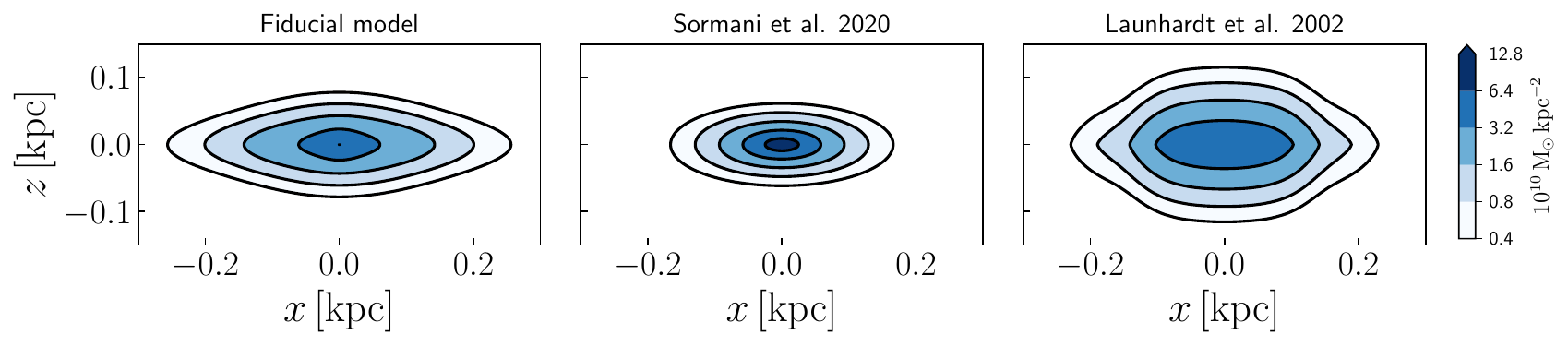}
    \caption{Surface density of our fiducial model compared to the best-fitting model of \citet{Launhardt2002} and model 3 of \citet{Sormani2020a}. The NSC is not included in this figure.}
    \label{fig:compare}
\end{figure*}

\begin{figure*}
	\includegraphics[width=\textwidth]{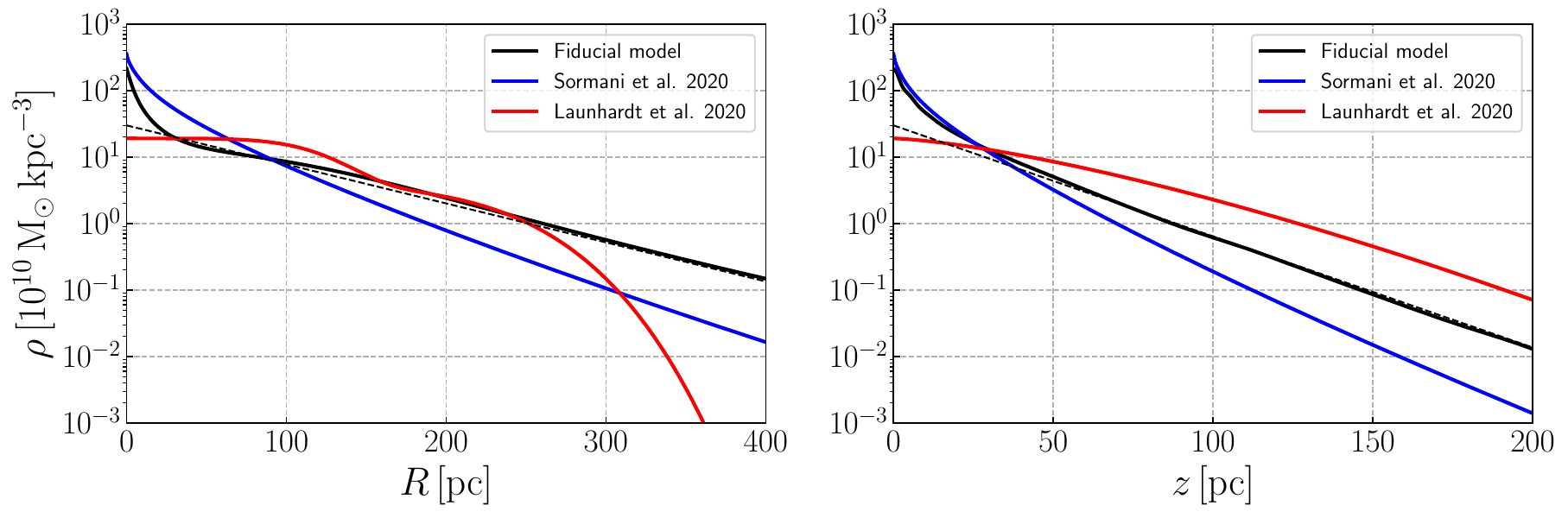}
    \caption{Density profiles of our fiducial model compared to the best-fitting model of \citet{Launhardt2002} and model 3 of \citet{Sormani2020a}. \emph{Left}: radial profiles along the plane $z=0$. \emph{Right}: vertical profiles along the axis $R=0$. The black dashed lines are exponential with scale-lengths $74\pc$ (left panel) and $26 \pc$ (right panel) respectively. The NSC is not included in this figure.}
    \label{fig:1Dcompare}
\end{figure*}

The radial scale-length according to the marginalised distributions in Figure~\ref{fig:mcmc} is $R_{\rm disc} = 88.6^{+9.2}_{-6.9} \pc$ and the vertical scale-height is $H_{\rm disc}=28.4^{+5.5}_{-5.5} \pc$. At first sight, these appear to be smaller than previous reported estimates (see Section~\ref{sec:structure}).
However, this discrepancy appears to be largely related to the different way in which these scale-lengths are defined. Figure~\ref{fig:1Dcompare} compares the radial and vertical profile of our fiducial NSD model to those of the model from \citet{Launhardt2002} and of model 3 from \citet{Sormani2020a}. The latter is obtained by deprojecting the S{\'e}rsic fit from \citet{Gallego-Cano2020} and so it automatically incorporates their scale-lengths. Figure~\ref{fig:1Dcompare} shows that the scale-lengths of our model are not smaller than those of previous models - if anything, they are slightly larger. Our fiducial model is similar to the model of \citet{Launhardt2002} outside the innermost 30pc, particularly in the plane $z=0$. This agreement is remarkable considering that our fitting procedure did not use any photometric information, and the scale-lengths are purely determined from the normalised kinematics distributions. The radial profile of the \citet{Sormani2020a} model (and therefore of \citealt{Gallego-Cano2020}) has somewhat shorter scale-lengths than our model, both vertically and radially in the plane $z=0$. Figure~\ref{fig:compare} compares the edge-on surface density of the three models, and shows that the aspect of our fiducial model is somewhere in between the \citet{Launhardt2002} and the \citet{Sormani2020a}/\citet{Gallego-Cano2020} models.

\subsection{Velocity dispersion} \label{sec:dispersion}

The top panel in Figure~\ref{fig:mom1} shows the velocity dispersions of the fiducial NSD model in the plane $z=0$. The dispersions decrease with radius. The fact that the total observed dispersion in the KMOS sample instead increases with radius (Fig.~12 in \citealt{Schultheis2021}) is explained in our model by increasing contamination of the Bar.

\citet{Sormani2020a} hypothesised based on their Jeans modelling that the NSD might be vertically biased, i.e.\ that it might have $\sigma_R /\sigma_z<1$. Such a property would not be obvious to explain in terms of secular heating mechanisms (see discussion in Sect.~5.2 of \citealt{Sormani2020a}). Here we find that our fiducial NSD stellar model is indeed vertically biased, but only at $R<30\pc$ (see bottom panel of Figure~\ref{fig:mom1} and Figure~\ref{fig:mom2}). This suggests that perhaps only the very central part of the NSD is vertically biased, perhaps because it contains a separate, more spheroidal component. In the model of \citet{Sormani2020a} the $\sigma_R /\sigma_z<1$ was needed to explain a drop in the second moment of the line-of-sight velocity dispersion at $|l|<0.5^\circ$. Since \citet{Sormani2020a} assume that the anisotropy parameter $b =\sigma_R^2 /\sigma_z^2$ is constant with radius, a common simplyfing assumption in Jeans modelling \citep{Cappellari2008}, their entire model is vertically biased, but one could modify their models to incorporate a radial-dependent $b$ so that the Jeans models would be biased only in central parts while still reproducing the data satisfactorily. However, we have experimented with manually changing the parameters of our models and we found plausible models that are not vertically biased. We conclude that while our model hints again at the possibility of a vertically biased disc, particularly in the central parts, evidence is not strong. 

Note that if the NSD is vertically biased only in the central part, this probably rules out the explanation given in \citet{Sormani2020a} that large vertical oscillations
are already imprinted at stellar birth, because stars are currently forming at the outer edge of the NSD, and so we would expect the disc to be biased also in the outer parts.

\subsection{Rotation of the NSD} \label{sec:rotation}

\begin{figure}
	\includegraphics[width=\columnwidth]{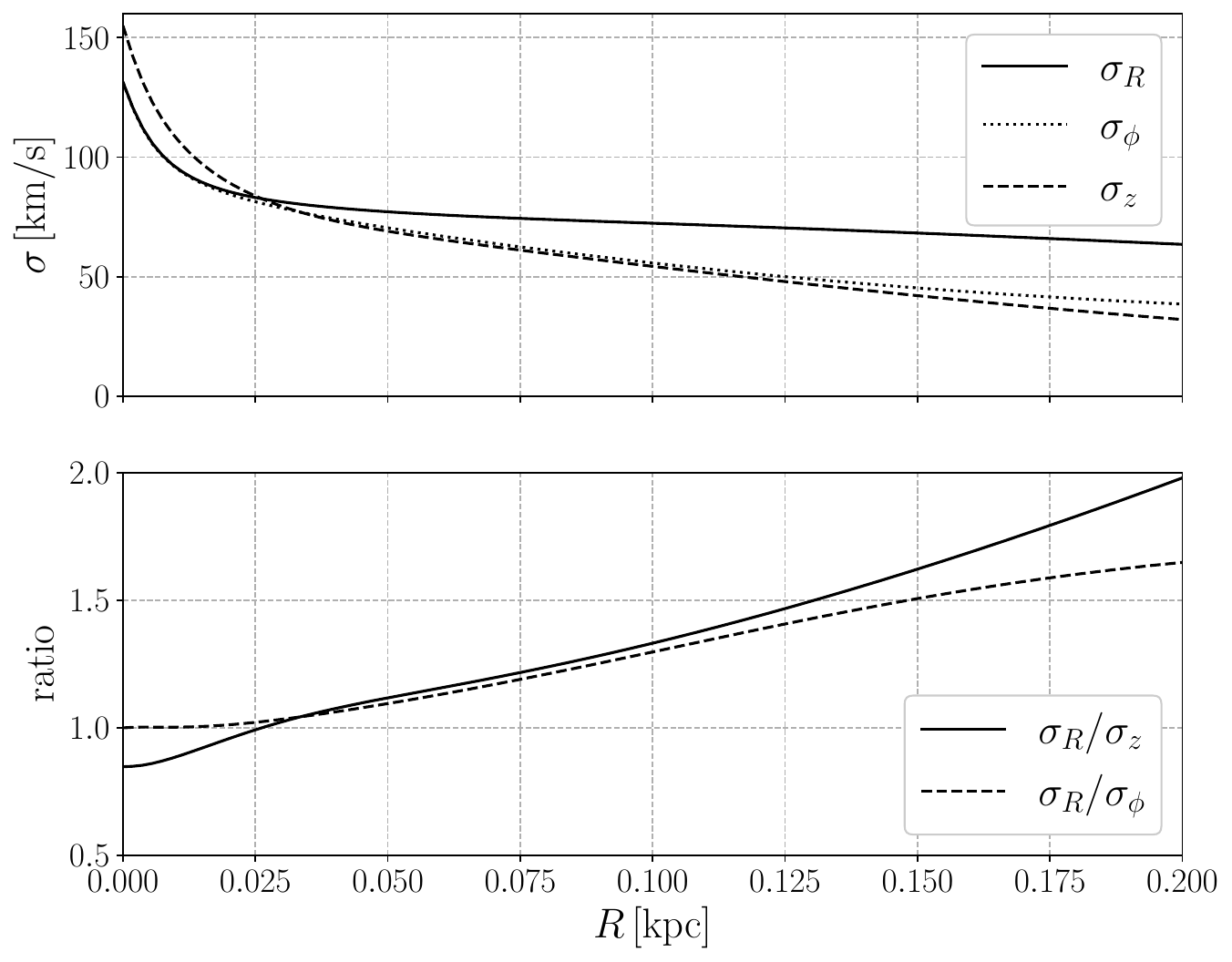}
    \caption{\emph{Top panel}: velocity dispersions of the NSD fiducial model in the plane $z=0$. \emph{Bottom panel}: ratios of the  velocity dispersions shown in the top panel. Note that these ratios are not constant, and that the NSD is vertically biased ($\sigma_R/\sigma_z<1$) at $R<30\pc$.}
    \label{fig:mom1}
\end{figure}

\begin{figure}
	\includegraphics[width=\columnwidth]{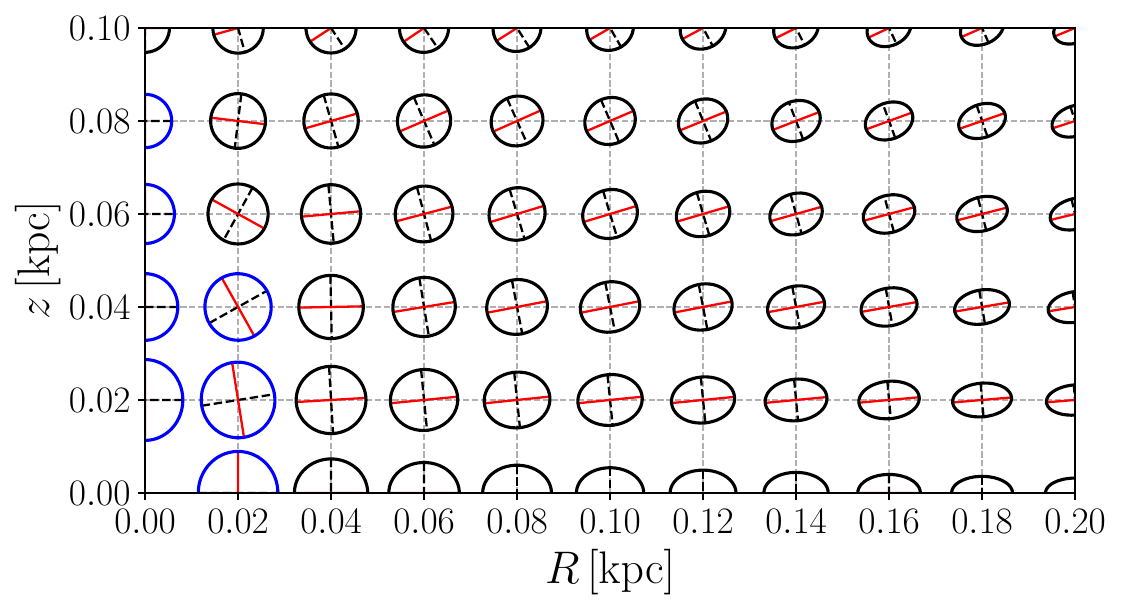}
    \caption{Representation of the velocity ellipsoids of the NSD fiducial model. Red is the major axis of the ellipse, while black dashed is the minor axis. The length of the red axes of each ellipse is proportional to the largest eigenvalues of the velocity ellipsoid at the centre of the ellipse. Blue indicates the ellipsoids that are vertically biased (i.e., the angle between the major axis and the $z$ axis is less than $45^\circ$).}
    \label{fig:mom2}
\end{figure}

\begin{figure}
	\includegraphics[width=\columnwidth]{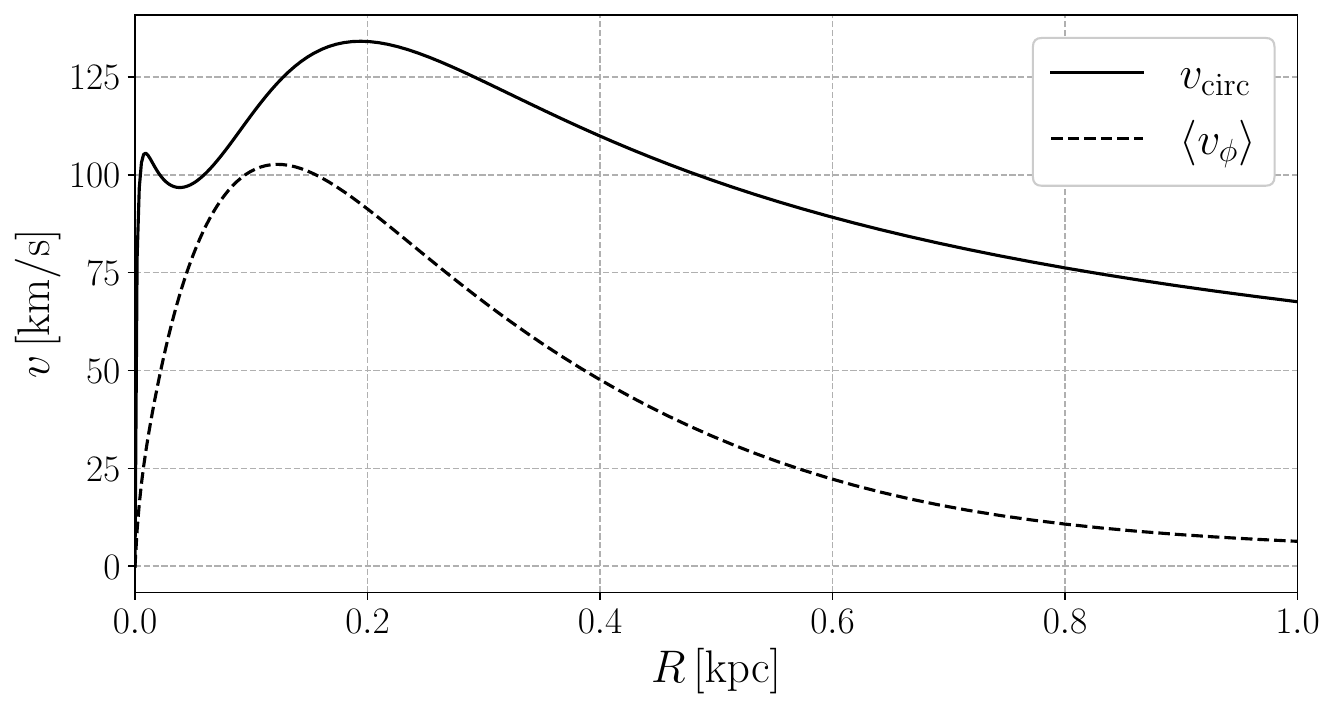}
    \caption{\emph{Solid line}: circular velocity curve $v_{\rm circ}=\sqrt{R \di \Phi/\di R}$ in the gravitational field generated by the sum of the NSC (Sect.~\ref{sec:NSC}) and our fiducial NSD model (Sect.~\ref{sec:fiducial}). The peak at $R\ll0.1\kpc$ is due to the NSC. \emph{Dashed line}: average azimuthal velocity $\langle v_\phi \rangle$ of NSD stars in our fiducial model. Both curves are in the plane $z=0$. The azimuthal velocity $\langle v_\phi \rangle$ is lower than $v_{\rm circ}$ because of asymmetric drift.}
    \label{fig:vcirc}
\end{figure}

The mean rotation velocity of the stars in the NSD (black dashed line in Figure~\ref{fig:vcirc}) is significantly lower than the rotation curve calculated from the gravitational potential (black solid in Figure~\ref{fig:vcirc}). This is a phenomenon called asymmetric drift \citep[e.g.][]{Binney2008}, and occurs when the velocity dispersion of the stars is non-negligible compared to their rotation velocity. In the case of the NSD the two are of the same order (compare Figures~\ref{fig:mom1} and \ref{fig:vcirc}), so this effect must be taken into account when estimating the gravitational field of the NSD from the rotation of the stars.

While in our model we average over the KMOS sample, in reality the asymmetric drift likely depends on the metallicity/age of the stars. \citet{Schultheis2021} find that metal-rich stars rotate faster and have lower velocity dispersion than metal-poor stars (see their Figures 10 and 13), and thus they display a smaller asymmetric drift. It is currently unclear whether this is because the metal-poor stars have a different origin than the metal-rich stars \citep{Schultheis2021}, or because the metal-poor stars are simply older and their velocity dispersion has increased over time. Also, the \cite{Schultheis2021} result might be affected by Bar pollution.

\subsection{Formation of the NSD} \label{sec:formation}

The NSD parameters derived in this paper imply that the NSD does not extend beyond the CMZ. This is consistent with the inside-out growth scenario proposed by \cite{Bittner2020} for nuclear discs in nearby galaxies. In this scenario, the NSD is formed from a series of ``CMZs'' ring-like structures that grow in radius over time.

The gravitational potential generated by the sum of our fiducial NSD model (Section \ref{sec:fiducial}), the \cite{Chatzopoulos2015b} NSC (Equation \ref{eq:NSC}) and the adopted P17 Bar model (Section \ref{sec:background}) has a single Inner Lindblad Resonance (ILR) placed at $R_{\rm ILR}=1.35\kpc$ in the epicyclic approximation. The location of the ILR is calculated assuming the nominal Bar pattern speed of $\Omega_{\rm p}=37.5\kms\kpc^{-1}$, and shifts to $R_{\rm ILR}=1.6\kpc$ and $R_{\rm ILR}=1.1\kpc$ if we take $\Omega_{\rm p}=35\kms\kpc^{-1}$ and $\Omega_{\rm p}=40\kms\kpc^{-1}$ respectively. This implies that the CMZ, as well as the stars currently forming inside it that will contribute to the the build-up of the NSD, are well inside the ILR. This is consistent with hydrodynamical simulations of gas flow in strongly barred potentials showing that the gaseous star-forming nuclear rings (corresponding to the CMZ in the Milky Way) form at a radius that, albeit correlated with the $R_{\rm ILR}$, it is smaller than the latter by a factor of several (see for example \citealt{Bissantz2003} and Fig.~6 in \citealt{Sormani2015a}).

It is likely that $R_{\rm ILR}$ was smaller in the past. As the Galaxy evolves over secular timescales, the ILR shifts outwards because of two effects: (1) The mass of the NSD grows over time. The increased central mass concentration changes the rotation curve and shifts the ILR outwards. (2) The Bar pattern speed steadily decreases with time, shifting the ILR outwards \citep{Debattista2000, Chiba2021a}. Since as mentioned above the size of the CMZ star-forming ring correlates with the radius of the ILR (see also references in Section \ref{sec:formation_intro}), this would also be consistent with the NSD being formed inside-out.

In conclusion, our recovered parameters of the NSD are fully consistent with the inside-out formation scenario. There are good prospects that we will be able to test this scenario in the near future. Indeed, the inside-out formation should leave clear signatures in the metallicities ($[\rm M/H]$), abundances ($[\alpha/\rm Fe]$) and ages of stars as a function of radius \citep{Bittner2020}. Spectroscopic observations from next-generation instruments such as CRIRES+ and MOONS (VLT) will allow us to measure these quantities for a statistically sufficient number of stars in the NSD.

\subsection{Other limitations of the model} \label{sec:extinction}

Extinction in the Galactic Centre is extremely inhomogeneous and clumpy (see for example \citealt{Schodel2014b}). Indeed, the Galactic Centre contains several compact dark clouds with extreme extinction \citep[e.g.][]{Henshaw2019,Zoccali2021,Nogueras-Lara2021a}. These can give rise to variations of extinction on scales smaller than the resolution of the \cite{Schultheis2014} map that we used in Section~\ref{sec:selection}, which has a resolution of $6'\times6'$. Thus it is possible that a small number of stars in the back side of the NSD are obscured in a way that is not captured by our modelling. This effect was modelled in the proper motion distribution of the NSC by \citet{Chatzopoulos2015a}. Modelling this effect for the NSD will require higher-resolution extinction maps and a larger data sample. Deeper data, that reach the Red Clump kinematically, will also provide useful information.

Our modelling assumes that the giant stars that constitute $>99\%$ of our sample are representative of the bulk of the stars that make up most of the NSD mass and that generate most of its gravitational potential. Most of stars in the NSD appear to be old \citep[e.g.][]{Minniti2016,ContrerasRamos2018,Nogueras-Lara2021c}. Using the NSD star formation history of \citet{Nogueras-Lara2020b} we roughly estimate that the contamination by young red giants (of age less than 1 Gyr) in our sample is not higher than about 10\%. Thus, our assumptions that stars in our sample trace the bulk of the mass seems plausible. However, if there are strong variations in the distributions of stars among stellar populations with different ages, our assumption might lead to biases. 

\section{Conclusions} \label{sec:conclusion}

We have constructed axisymmetric self-consistent dynamical models of the Milky Way's Nuclear Stellar Disc in which the distribution function is an analytic function of the action variables. We fitted them to the normalised kinematic distributions (line-of-sight and proper motions) of stars in the survey of \citet{Fritz2021} cross-matched with the VIRAC2 catalogue. Our fitting procedure is purely based on the kinematics and uses no photometric information. We found the following results:
\begin{enumerate}
    \item The mass of the NSD is $M_{\rm NSD} = 10.5^{+1.1}_{-1.0} \times10^8 \,\Msun$, consistent with previous independent photometric determinations (Sections~\ref{sec:around} and \ref{sec:mass}).
    \item The NSD has approximately exponential radial scale-length and vertical scale-height of $R_{\rm disc} = 88.6^{+9.2}_{-6.9} \pc$ and $H_{\rm disc}=28.4^{+5.5}_{-5.5} \pc$ respectively. 
    The density profiles are similar to those obtained with different methods by previous photometric models (see Sections~\ref{sec:around} and \ref{sec:scaleheight} and Figures \ref{fig:compare} and \ref{fig:1Dcompare}).
    \item The velocity dispersion of stars in the NSD is $\sigma \simeq 70\kms$ and decreases with radius (see Section~\ref{sec:dispersion} and Figure~\ref{fig:mom1}).
    \item The assumption of axisymmetry gives an adequate representation of the data. There is no obvious signature for the presence of a nuclear bar (Section~\ref{sec:fiducial}).
    \item Contamination from the Galactic Bar in the survey of \citet{Fritz2021} is significant. We provide a table with the level of contamination for each field according to our fiducial model (see Section~\ref{sec:fiducial}, Table~\ref{tab:2} and Figure~\ref{fig:ratio}).
    \item Our models provide the best constraints to date on the rotation curve in the innermost few hundred parsecs of the Milky Way (see Section~\ref{sec:rotation} and Figure~\ref{fig:vcirc}).
    \item Although it cannot be ruled out that the NSD is vertically biased as suggested in \cite{Sormani2020a}, especially in the inner parts, evidence for it is not strong (see Section~\ref{sec:dispersion} and Figure~\ref{fig:mom2}).
\end{enumerate}
Our fiducial model of the NSD provides the full 6D distribution function in phase space $(\bfx,\bfv)$ and is made publicly available as part of the software \textsc{Agama}. This model can be used to generate predictions for future surveys and other applications. Worthwhile directions for future investigations include modelling the NSC and the NSD simultaneously, and extending the distribution function of the models to include explicit dependencies on metallicity and age with the aim to test the inside-out formation scenario \citep{Sanders2015,Bittner2020}.

\section*{Acknowledgements}

We thank the referee for constructive comments that improved the paper, and Dimitri Gadotti and Zhi Li for useful comments on an earlier draft of this paper. MCS acknowledges financial support from the European Research Council via the ERC Synergy Grant ``ECOGAL – Understanding our Galactic ecosystem: from the disk
of the Milky Way to the formation sites of stars and planets'' (grant 855130). JLS thanks the support of the Royal Society (URF\textbackslash R1\textbackslash191555). EV acknowledges support from STFC via the consolidated grant to the Institute of Astronomy. DM acknowledges support from project CATA FB210003. AMB acknowledges funding from the EU Horizon 2020 research and innovation programme under the Marie Sk\l{}odowska-Curie grant agreement No 895174. JM acknowledges support from the UK Science and Technology Facilities Council under grant number ST/S000488/I. MCS, NN, FN-L and RSK gratefully acknowledge support by the Deutsche Forschungsgemeinschaft (DFG, German Research Foundation) – Project-ID 138713538 – SFB 881 (“The Milky Way System”, subprojects B1, B2, B8). FN-L acknowledges the sponsorship provided by the Federal Ministry for Education and Research of Germany through the Alexander von Humboldt Foundation. RS and BS acknowledge financial support from the State Agency for Research of the Spanish MCIU through the ``Center of Excellence Severo Ochoa'' award for the Instituto de Astrof\'isica de Andaluc\'ia (SEV-2017-0709) and financial support from national project PGC2018-095049-B-C21 (MCIU/AEI/FEDER, UE). RSK acknowledges support from the Heidelberg cluster of excellence EXC 2181 (Project-ID 390900948) ``STRUCTURES: A unifying approach to emergent phenomena in the physical world, mathematics, and complex data'' funded by the German Excellence Strategy. This work made use of computing resources provided by the state of Baden-W\"{u}rttemberg through bwHPC and the German Research Foundation (DFG) through grant INST 35/1134-1 FUGG. Data are stored at SDS@hd supported by the Ministry of Science, Research and the Arts Baden-W\"urttemberg (MWK) and DFG through grant INST 35/1314-1 FUGG.

\section*{Data Availability}

The code that reproduces our fiducial model is publicly available as one of the python examples included in the software package \textsc{Agama} (\url{https://github.com/GalacticDynamics-Oxford/Agama}). The spectroscopic survey of \citet{Fritz2021} is publicly available. The proper motions obtained by cross matching with the VIRAC2 reduction of VVV  will be published together with the second version of the VIRAC catalogue (Smith et al., in prep.).


\bibliographystyle{mnras}
\bibliography{bibliography}



\appendix

\section{Detailed comparison between the fiducial model and the data in each KMOS field} \label{sec:appendix_hist}

\begin{figure*}
	\includegraphics[width=\textwidth]{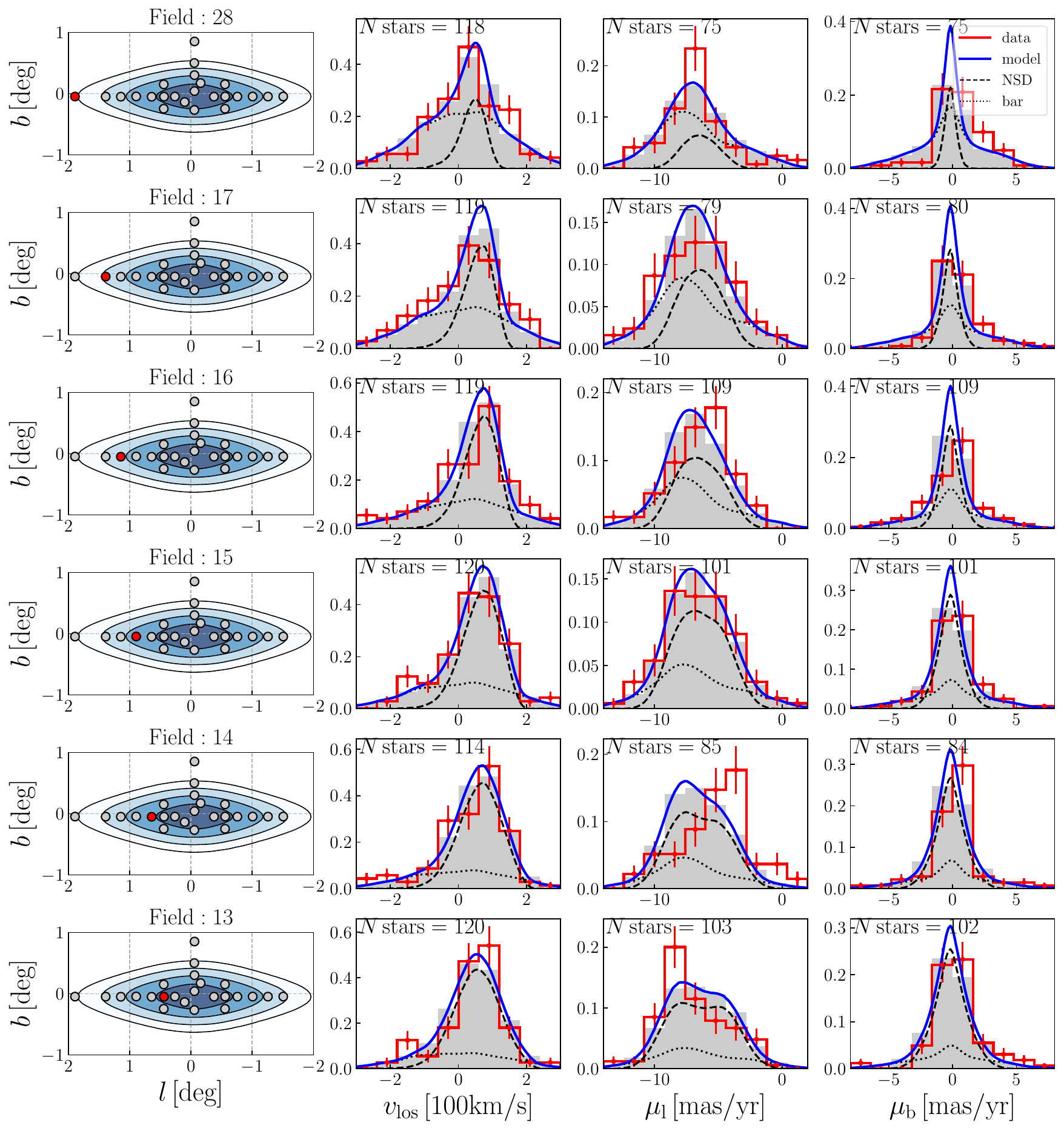}
    \caption{Detailed comparison between our fiducial model and the KMOS data cross matched to VIRAC2. Each row corresponds to an individual field. The three rightmost panels in each row show the normalised kinematic distributions corresponding to the field highlighted in red in the leftmost panel. \emph{Red solid}: histogram of the KMOS data (Section \ref{sec:data}). The error bars show the $\sqrt{N}$ shot noise. The numbers in the top-left of each panel indicate the total number of stars in the histogram. \emph{Black dashed:} the NSD contribution (Section \ref{sec:fnsd}). \emph{Black dotted:} the Bar/Disc contribution (Section \ref{sec:background}). \emph{Blue solid}: the sum of the two. \emph{Gray filled:} histogram of the model (NSD+Bar, blue line) binned in the same way as the data.}
    \label{fig:scm_hist_0}
\end{figure*}

\begin{figure*}
	\includegraphics[width=\textwidth]{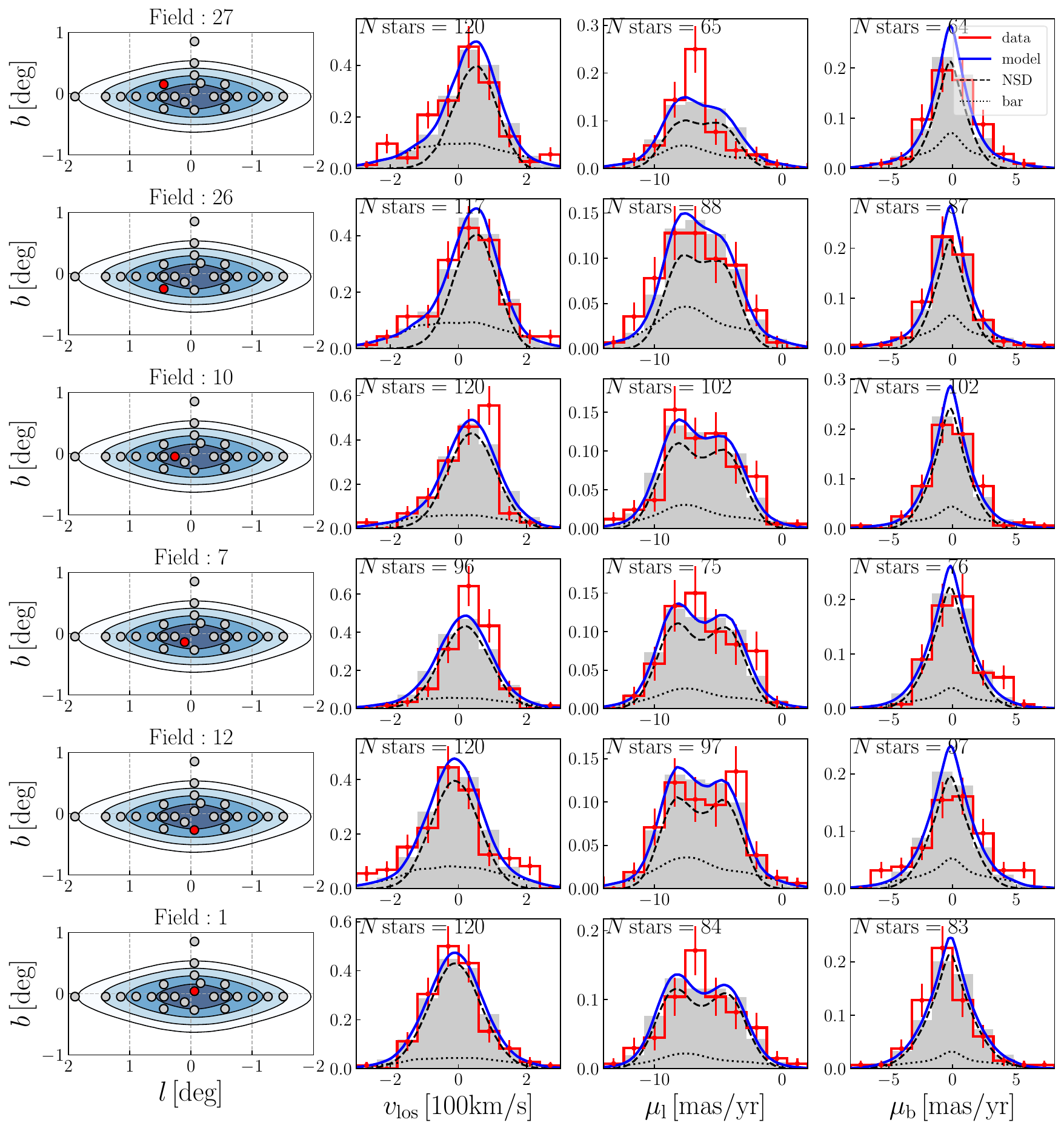}
    \caption{Same as Figure~\ref{fig:scm_hist_0} for 6 more KMOS fields.}
    \label{fig:scm_hist_1}
\end{figure*}

\begin{figure*}
	\includegraphics[width=\textwidth]{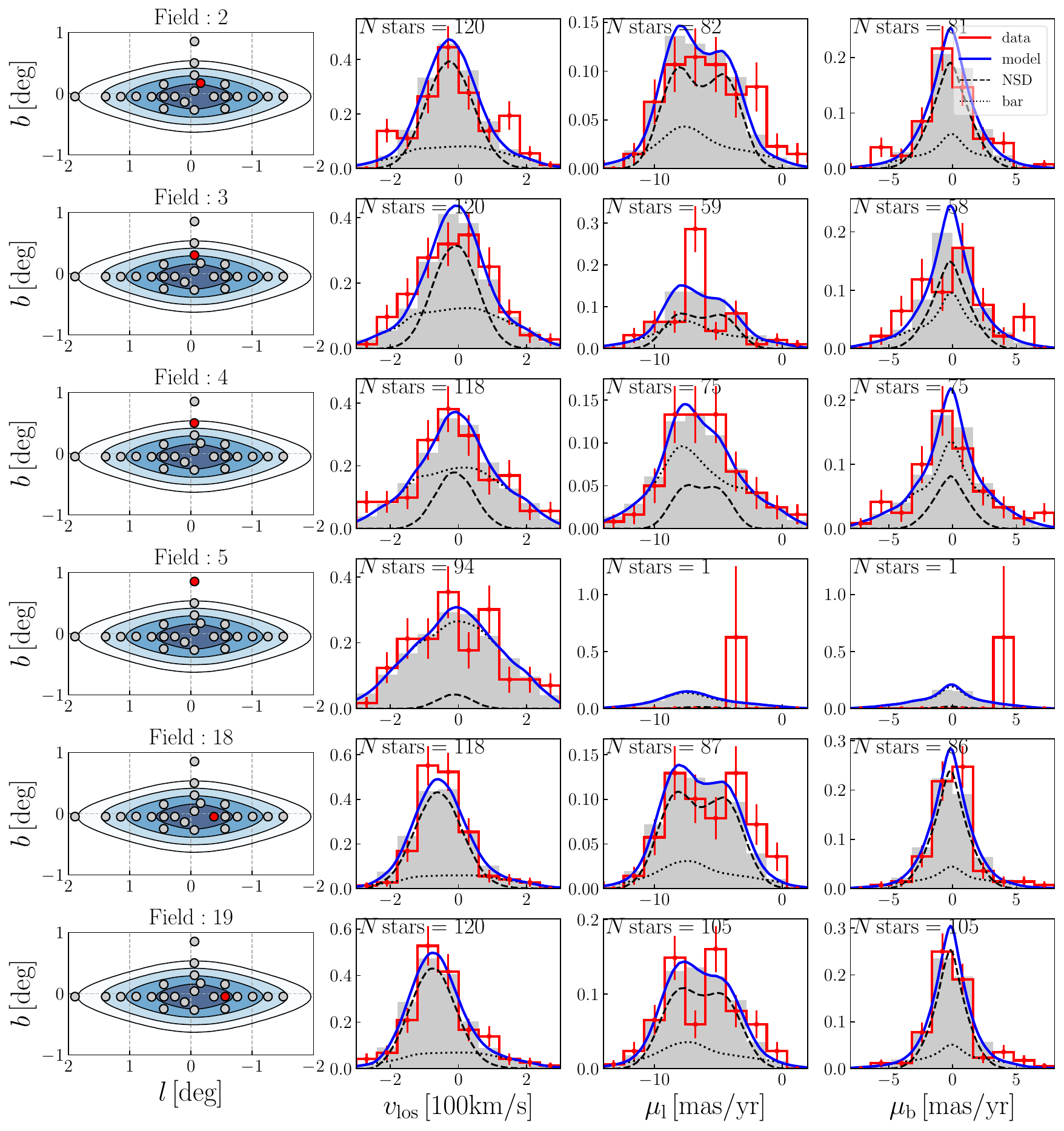}
    \caption{Same as Figure~\ref{fig:scm_hist_0} for 6 more KMOS fields.}
    \label{fig:scm_hist_2}
\end{figure*}

\begin{figure*}
	\includegraphics[width=\textwidth]{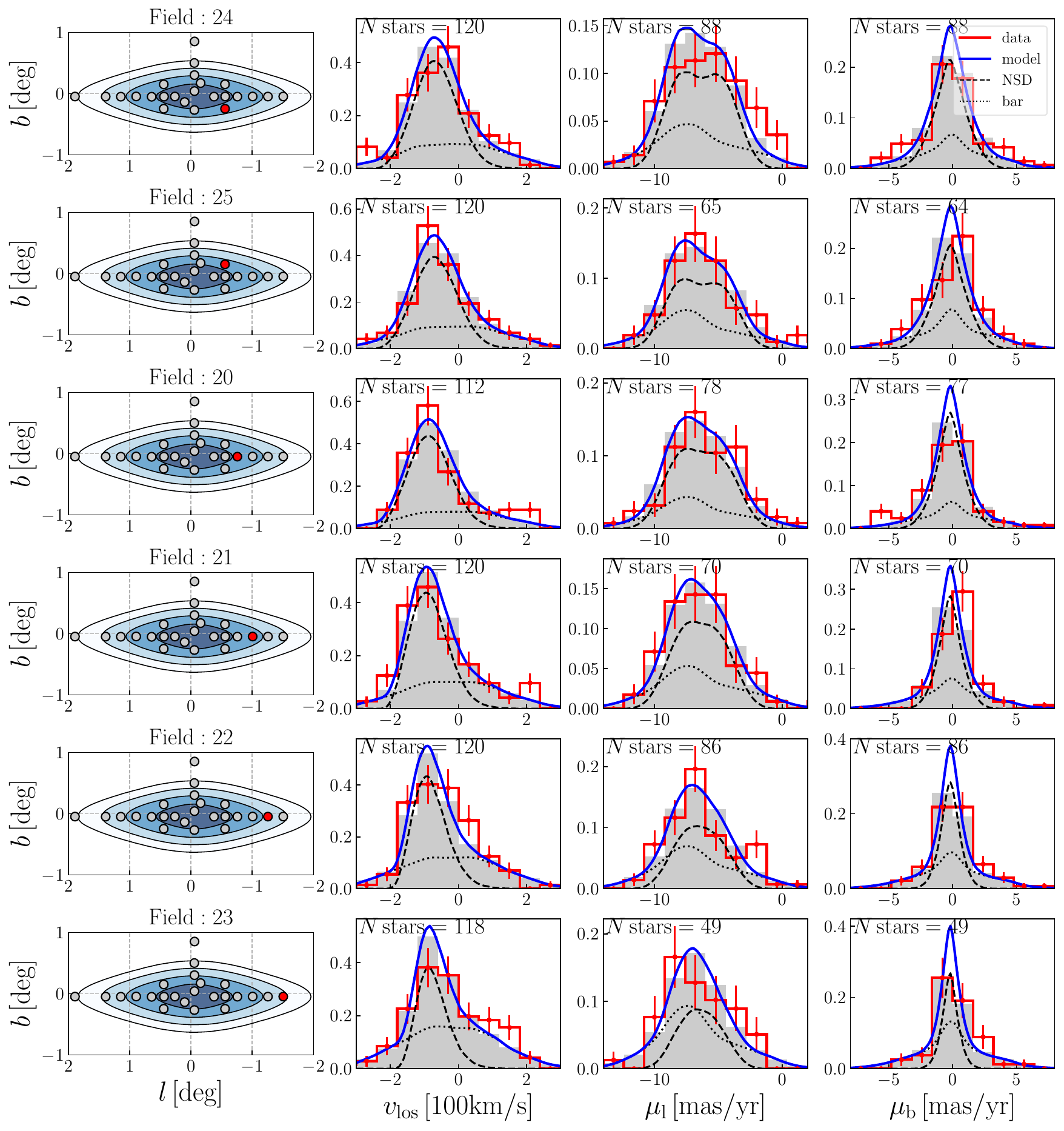}
    \caption{Same as Figure~\ref{fig:scm_hist_0} for 6 more KMOS fields.}
    \label{fig:scm_hist_3}
\end{figure*}

Figures~\ref{fig:scm_hist_0}-\ref{fig:scm_hist_3} show a detailed field-by-field comparison of our fiducial model (Table~\ref{tab:1}) and the data. Note that the histograms shown in these figures are only to facilitate comparison, since the fitting is done on a star-by-star basis and does not involve binning (see Section~\ref{sec:fitting}).

\section{Selection fractions for all fields} \label{sec:appendix_sel}

\begin{figure*}
	\includegraphics[width=\textwidth]{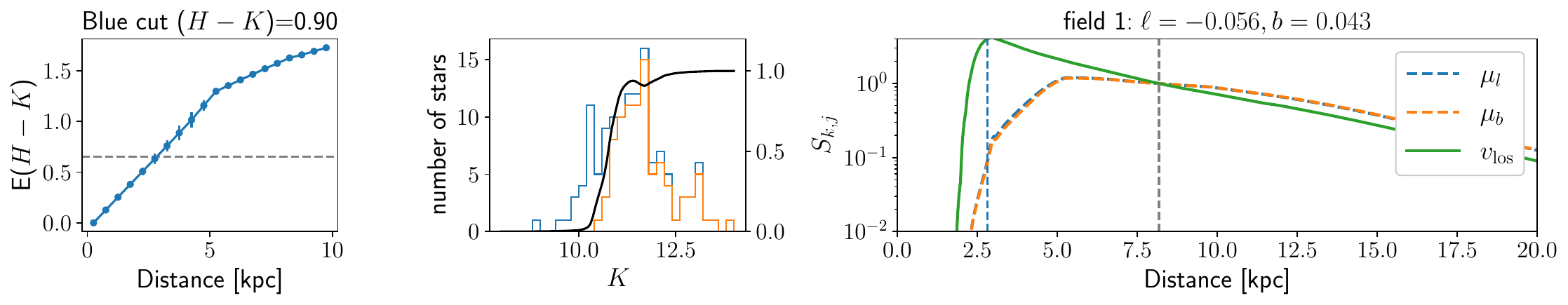}
	\includegraphics[width=\textwidth]{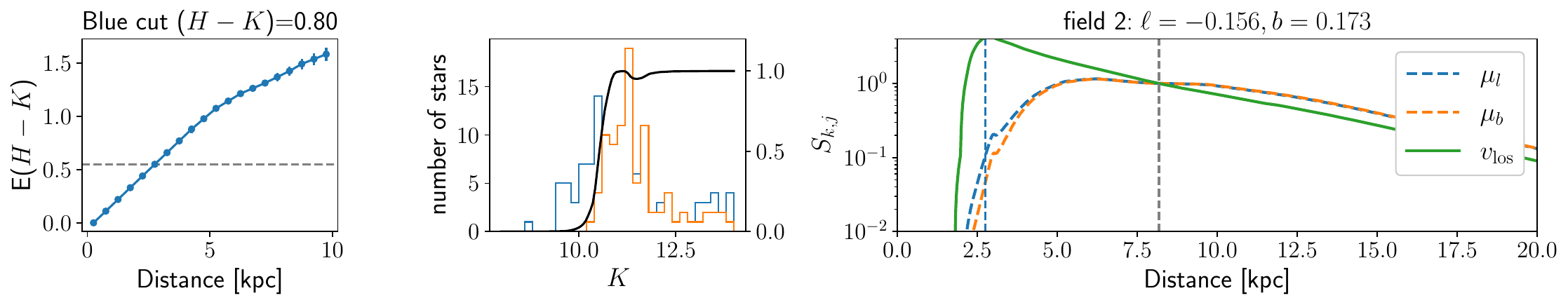}
	\includegraphics[width=\textwidth]{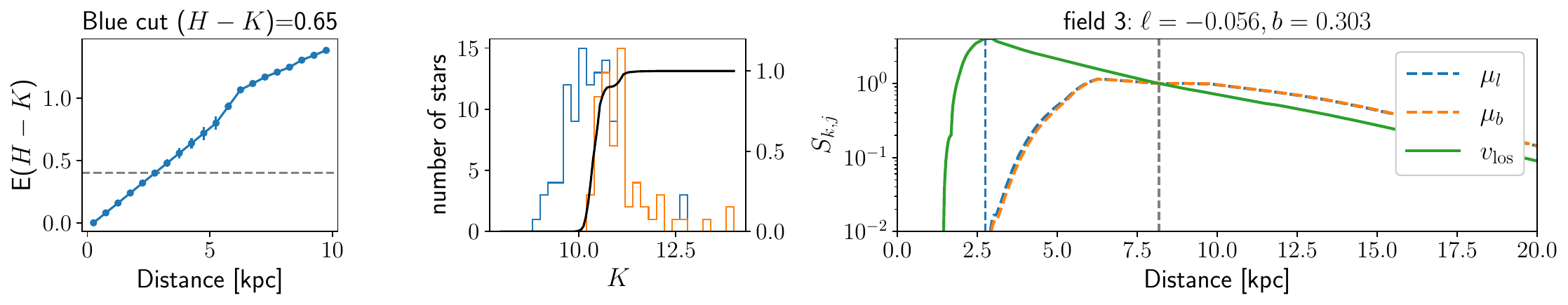}
	\includegraphics[width=\textwidth]{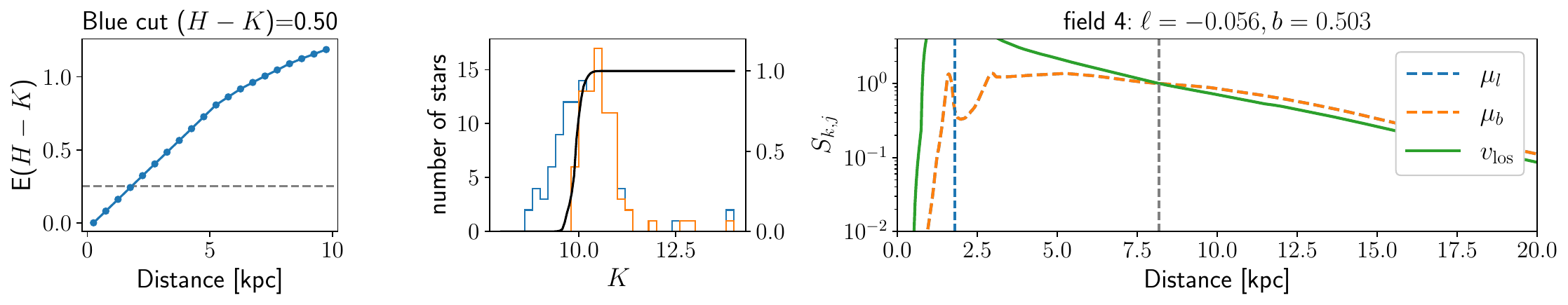}
	\includegraphics[width=\textwidth]{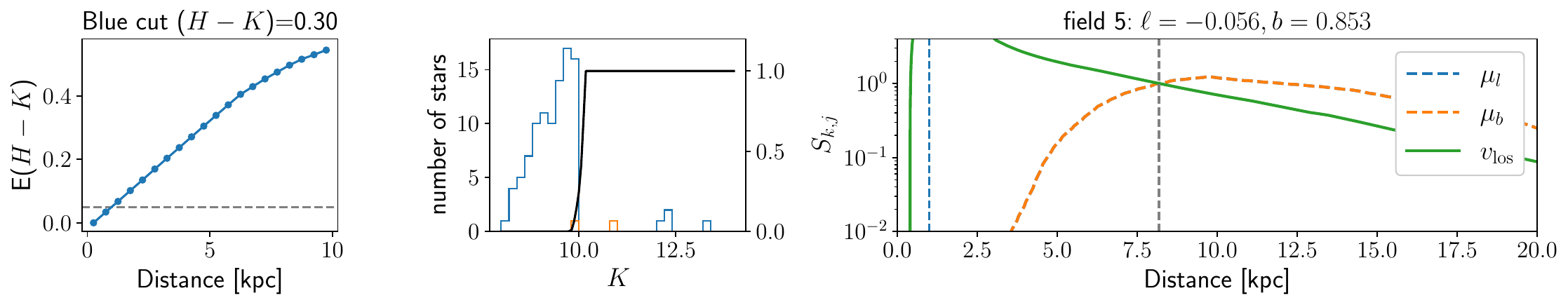}
	\includegraphics[width=\textwidth]{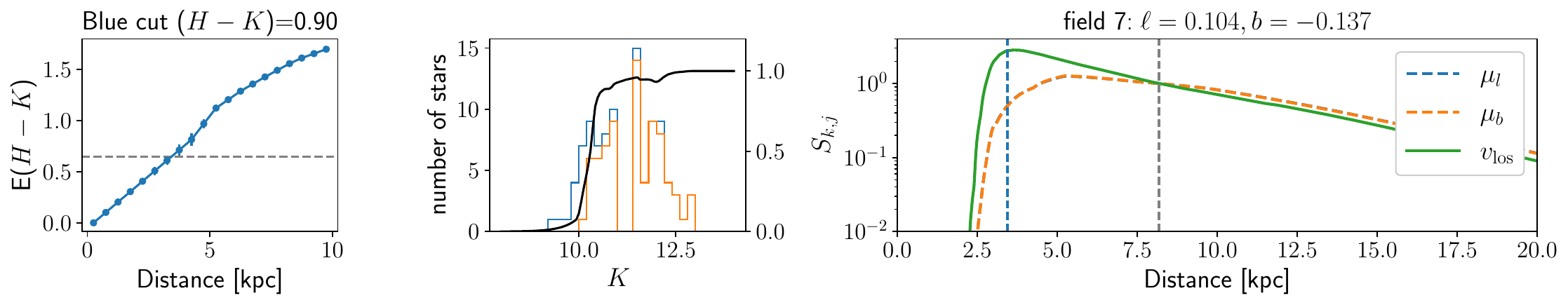}
    \caption{\emph{Left panels}: the colour excess $E(H-K)=(H-K) - (H-K)_0$ as a function of distance from the 3D maps of \citet{Schultheis2014}. The horizontal dashed line is $(H-K)_{\rm cut} - 0.25$, where 0.25 is the typical intrinsic colour of bright stars that constitute the majority of stars in our sample. The distance at which this line intersects the $E(H-K)$ gives a good idea of where the selection fraction shown in the right-hand panel drops. \emph{Middle panels}: histogram of all stars in apparent magnitude $K$ (blue) and only of stars with good proper motions ($
    \mu_{l,{\rm err}}<1 \masyr$, orange). The black full line is the ratio between the two (right axis). \emph{Right panels}: the selection fraction $S_{k,j}$. The gray dashed line indicates the distance of the Galactic Centre. The blue dashed line indicates where the drop should occur according to the intersection point in the left panel. For the central fields, the cut is at distance of around 3 kpc, consistent with \citet{Nogueras-Lara2021d}.}
    \label{fig:sel_01}
\end{figure*}

\begin{figure*}
	\includegraphics[width=\textwidth]{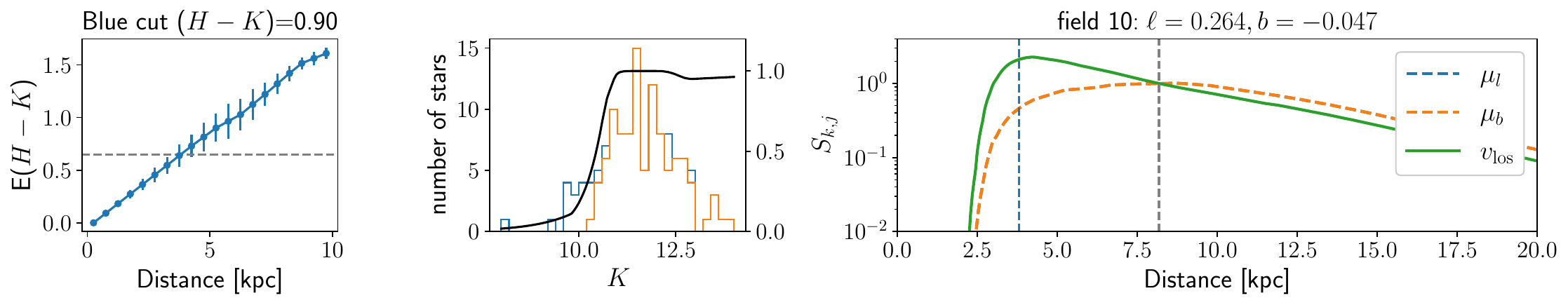}
	\includegraphics[width=\textwidth]{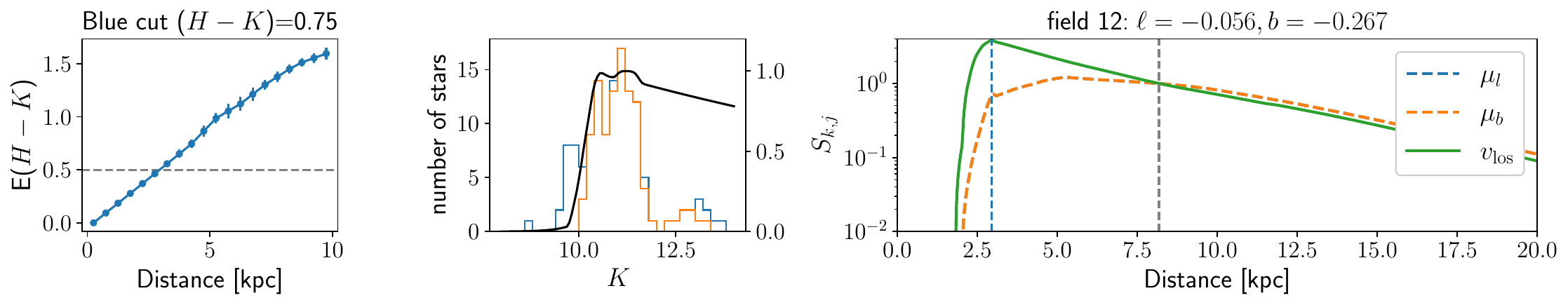}
	\includegraphics[width=\textwidth]{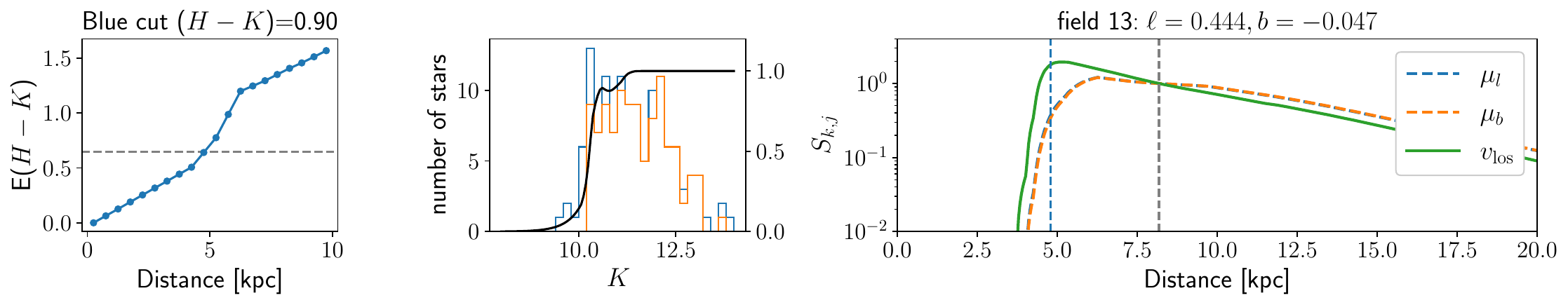}
	\includegraphics[width=\textwidth]{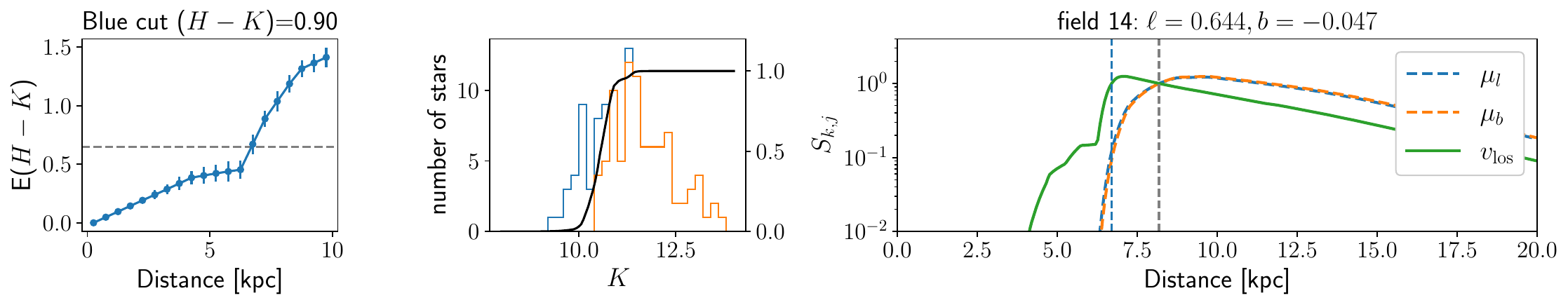}
	\includegraphics[width=\textwidth]{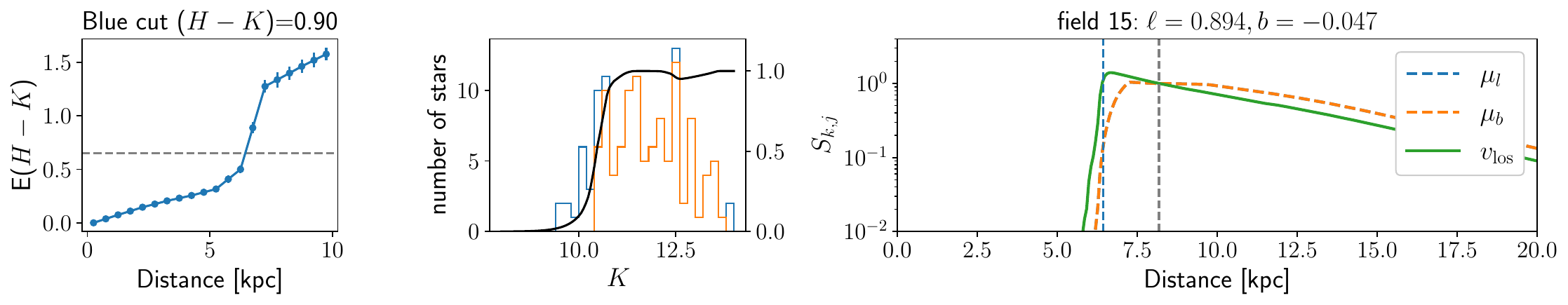}
	\includegraphics[width=\textwidth]{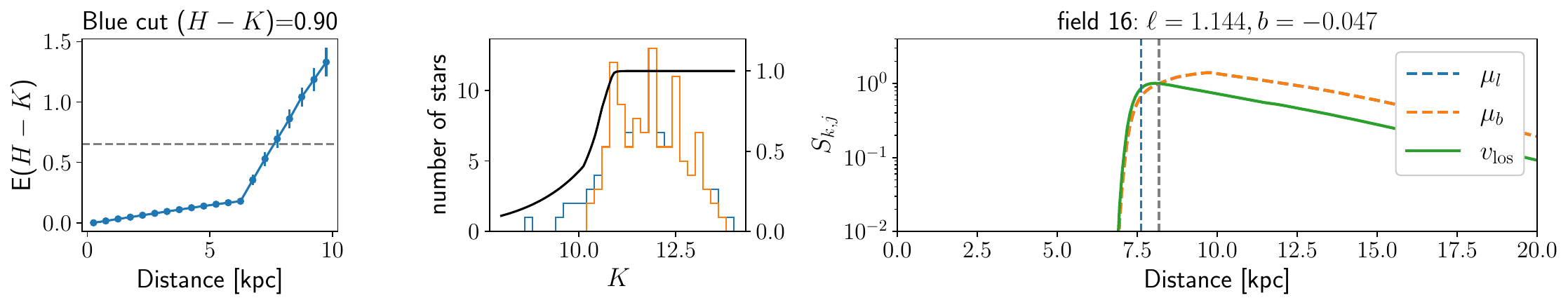}
    \caption{Same as Figure~\ref{fig:sel_01} for 6 more KMOS fields.}
    \label{fig:sel_02}
\end{figure*}

\begin{figure*}
	\includegraphics[width=\textwidth]{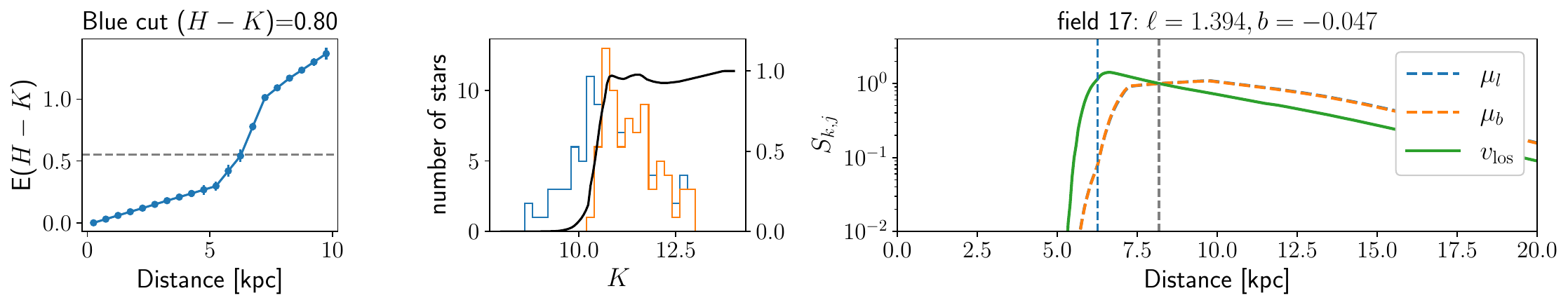}
	\includegraphics[width=\textwidth]{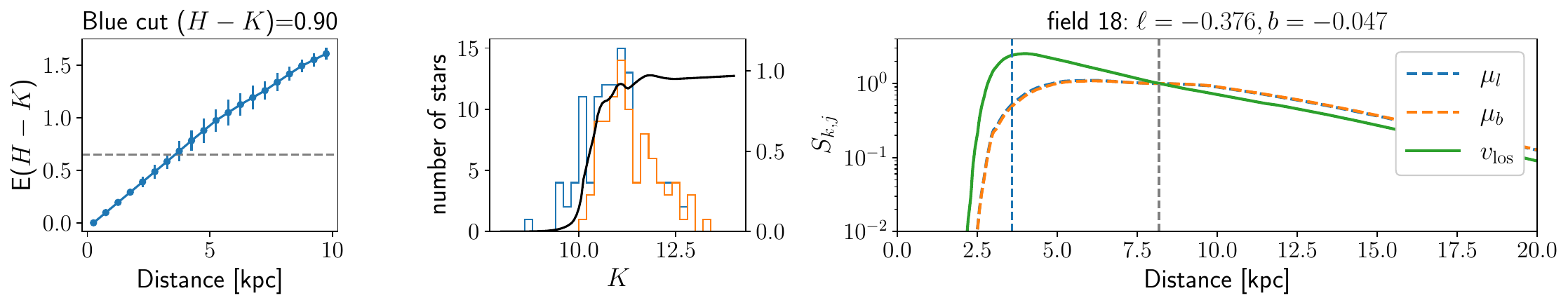}
	\includegraphics[width=\textwidth]{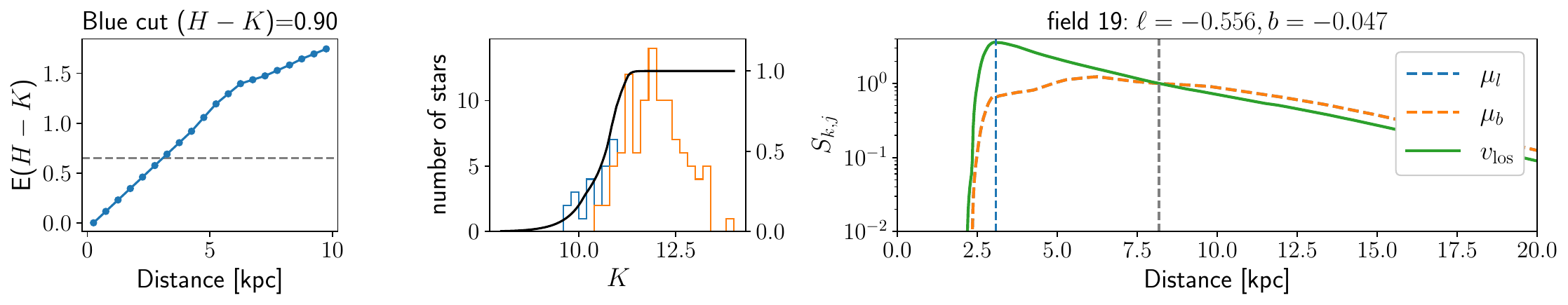}
	\includegraphics[width=\textwidth]{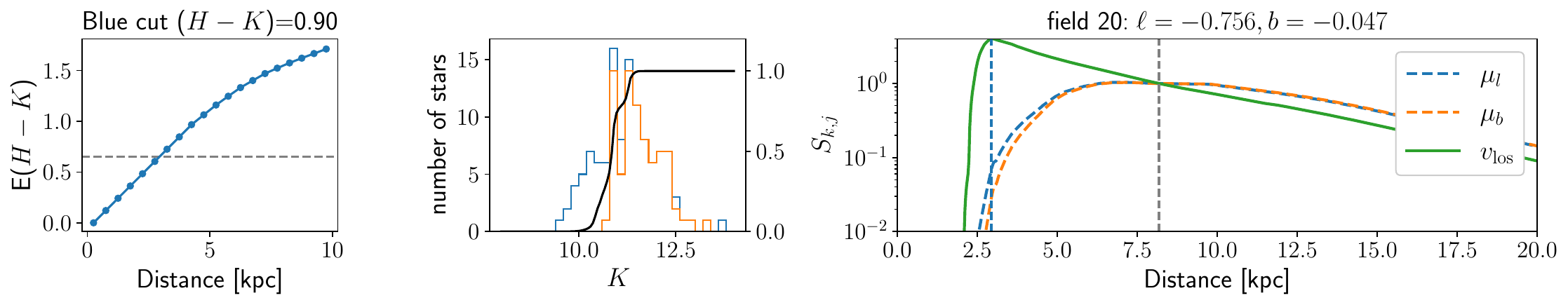}
	\includegraphics[width=\textwidth]{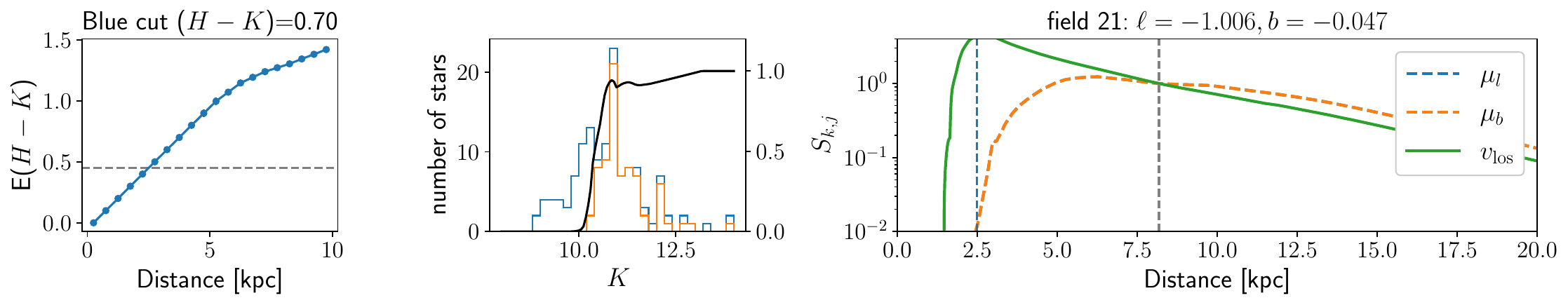}
	\includegraphics[width=\textwidth]{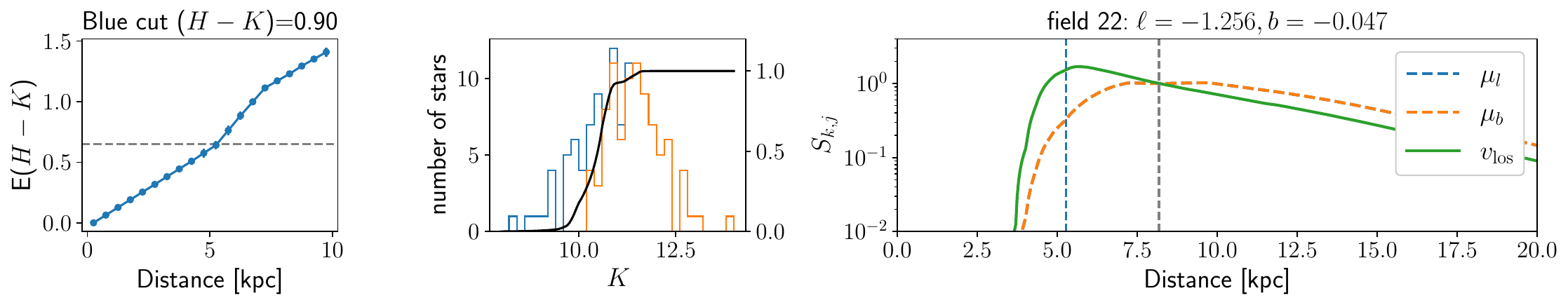}
    \caption{Same as Figure~\ref{fig:sel_01} for 6 more KMOS fields.}
    \label{fig:sel_03}
\end{figure*}

\begin{figure*}
	\includegraphics[width=\textwidth]{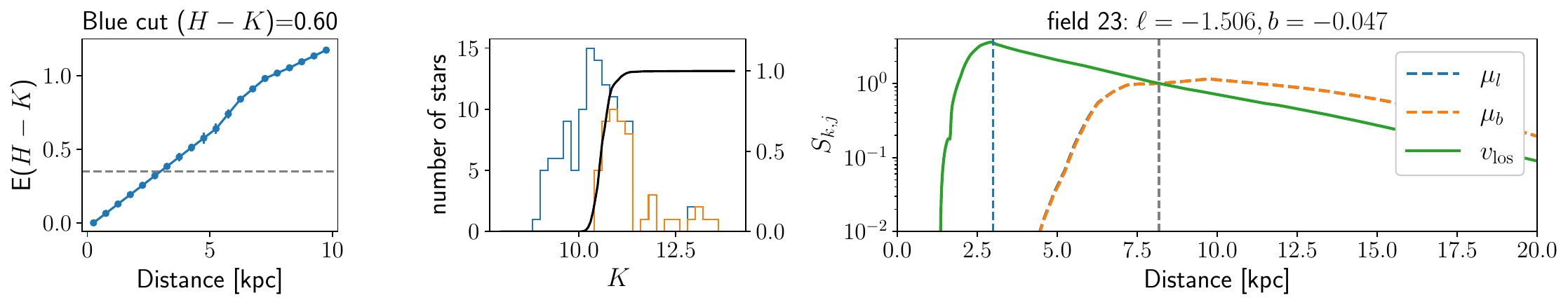}
	\includegraphics[width=\textwidth]{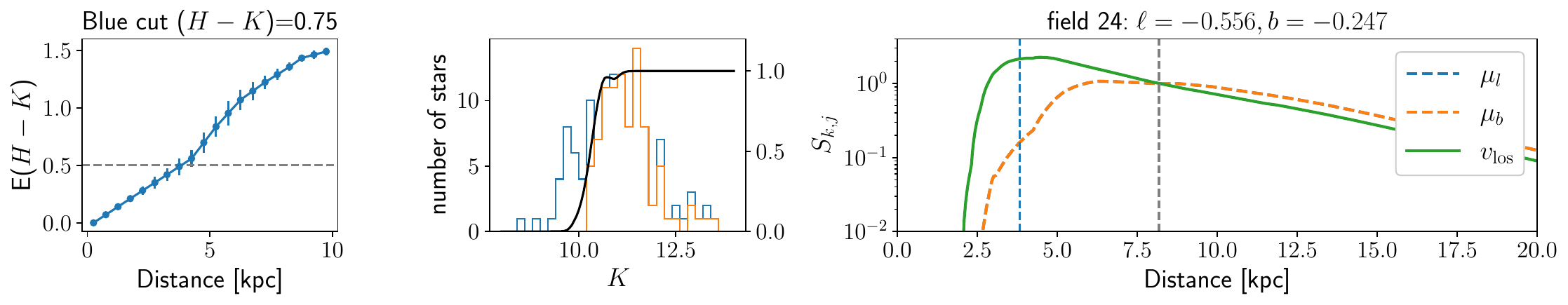}
	\includegraphics[width=\textwidth]{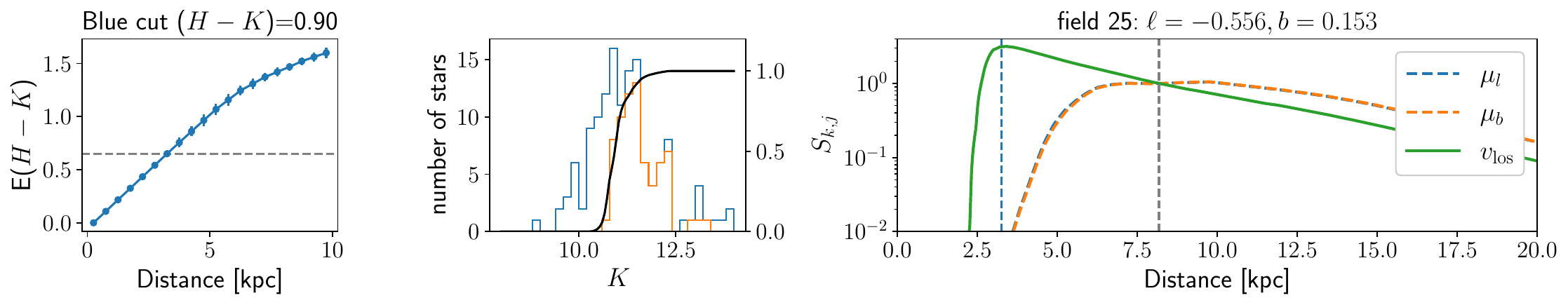}
	\includegraphics[width=\textwidth]{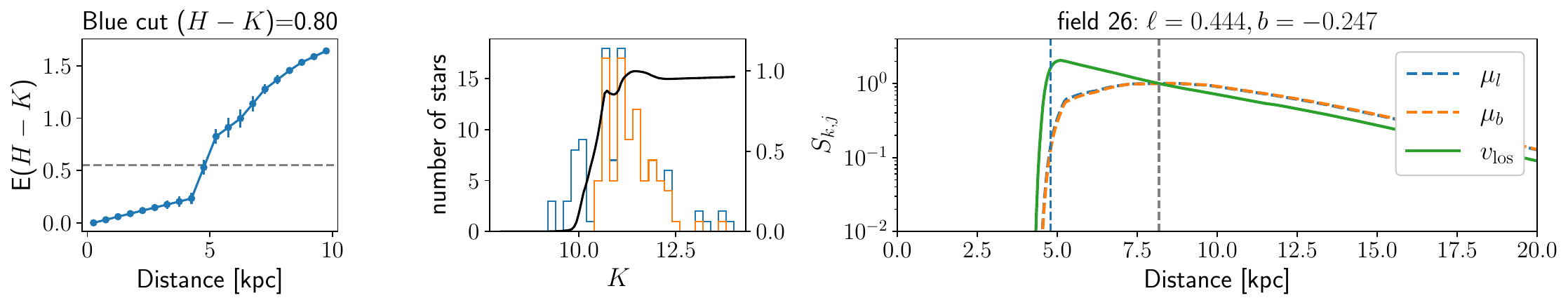}
	\includegraphics[width=\textwidth]{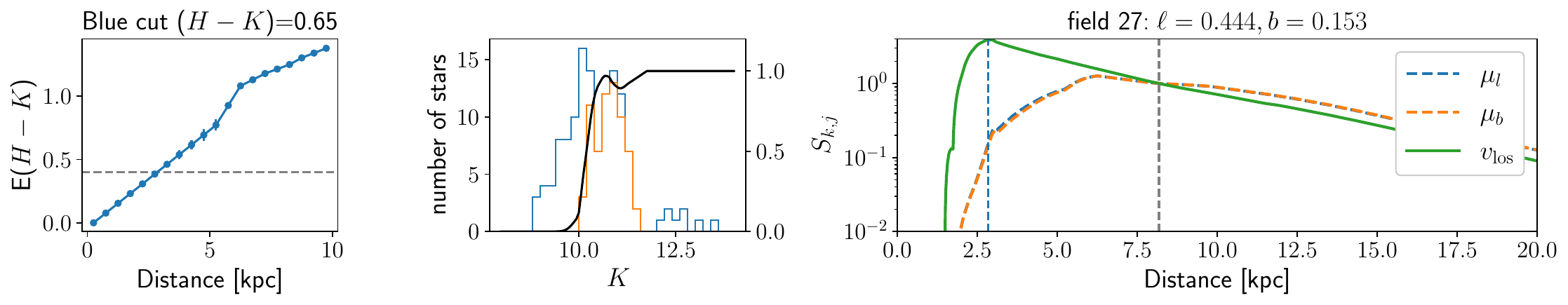}
	\includegraphics[width=\textwidth]{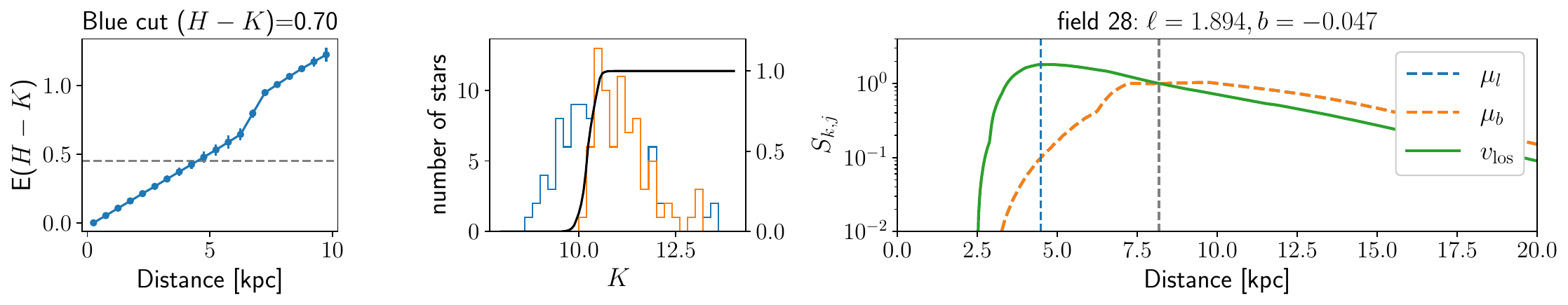}
    \caption{Same as Figure~\ref{fig:sel_01} for 6 more KMOS fields.}
    \label{fig:sel_04}
\end{figure*}

In the main text we have shown the selection fraction for one example field (Figure~\ref{fig:sel_main}). In this appendix we show the selection fraction for all the 24 fields (see Figures~\ref{fig:sel_01}-\ref{fig:sel_04}).

\section{Illustration of why the $\mu_l$ distribution of the Bar is skewed} \label{sec:appendix_skewed}

Here we give a brief explanation of the geometric effect that causes the Bar background $\mu_l$ distributions in Figures~\ref{fig:scm_hist_0}-\ref{fig:scm_hist_3} to be skewed and have a shoulder, using a toy model.

\begin{figure*}
	\includegraphics[width=\textwidth]{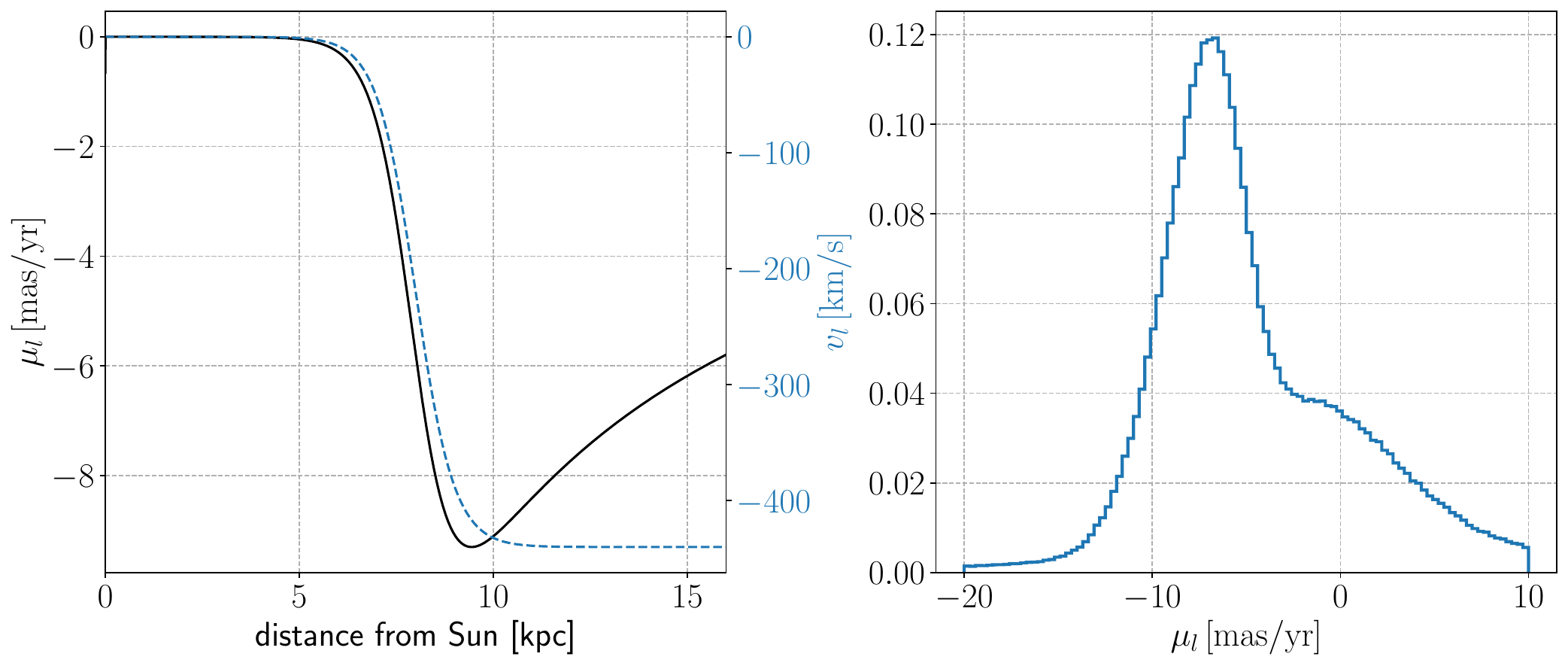}
    \caption{\emph{Left:} tangential velocity $v_l=\mu_l \times \text{distance}$ and proper motion $\mu_l$ along a line of sight at $l=0$ in the toy model described in Appendix~\ref{sec:appendix_skewed}. \emph{Right:} $\mu_l$ distribution of stars along this line of sight in the toy model assuming a random velocity dispersion of $100\kms$.}
    \label{fig:skewed}
\end{figure*}

Consider an axisymmetric disc of stars in circular motion distributed with uniform density inside the Solar circle ($R<8\kpc$). Assume that the rotation curve is simply $v_{\rm circ} = 220 \kms \times \tanh(R/\kpc) $ and that the Sun is on a purely circular orbit with $v=220\kms$. The left panel in Figure~\ref{fig:skewed} shows the tangential velocity $v_l$ and the proper motion $\mu_l$ along a line of sight that goes exactly through the Galactic Centre ($l=0$) in this simple toy model.

If we take stars along this line-of-sight, add a random velocity dispersion of $100\kms$ and plot the resulting distribution, we obtain the histogram in the right panel of Fig.~\ref{fig:skewed}. One can see that there is an accumulation of stars at lower values of $\mu_l$ that skews the distribution to the left. The result is qualitatively similar to the Bar background distributions in Figures~\ref{fig:scm_hist_0}-\ref{fig:scm_hist_3}. The reason for the skewness is that the proper motion is defined as $\mu_l=v_l/{\rm distance}$, so points that are at the symmetric positions with respect to the Galactic Centre do not have symmetric values of $\mu_l$, despite having symmetric values of $v_l$. This can be seen from the left panel in Figure~\ref{fig:skewed}. The effect becomes more and more important as one considers stars over a more extended region along the line-of-sight.

\bsp	
\label{lastpage}
\end{document}